\documentclass[12pt,compress,nosort]{JHEP3}
\usepackage{exscale,latexsym}
\usepackage[dvips]{graphicx}
\usepackage{axodraw}
\usepackage{cite}

\def\Year{\expandafter\eatPrefix\the\year}
\newcount\hours \newcount\minutes
\def\monthname{\ifcase\month\or
January\or February\or March\or April\or May\or June\or July\or
August\or September\or October\or November\or December\fi}\def\shortmonthname{\ifcase\month\orx
Jan\or Feb\or Mar\or Apr\or May\or Jun\or Jul\or
Aug\or Sep\or Oct\or Nov\or Dec\fi}

\def\TimeStamp{\hours\the\time\divide\hours by60%
\minutes -\the\time\divide\minutes by60\multiply\minutes by60%
\advance\minutes by\the\time%
${\rm \shortmonthname}\cdot   \if\day<10{}0\fi\the\day\cdot   \the\year
\qquad\the\hours:\if\minutes<10{}0\fi\the\minutes$}

\newskip\humongous \humongous=0pt plus 1000pt minus 100pt
\def\caja{\mathsurround=0pt}
\def\eqalign#1{\,\vcenter{\openup1\jot \caja
       \ialign{\strut \hfil$\displaystyle{##}$&$
        \displaystyle{{}##}$\hfil\crcr#1\crcr}}\,}
\newif\ifdtup

\newcounter{eqnumber}[section]
\renewcommand{\theeqnumber}{\thesection.\arabic{eqnumber}}
\def\equn{\refstepcounter{eqnumber}
\eqno({\rm \theeqnumber})
}

\def\Fs#1#2{F^{{#1}}_{}}
\def\Fone{\Fs{\rm 1m}}
\def\Feasy{\Fs{{\rm 2m}\,e}}

\def\Fhard{\Fs{{\rm 2m}\,h}}

\def\Fthree{\Fs{\rm 3m}}

\def\cg{c_\Gamma}
\def\rg{r_\Gamma}

\def\Gr{{\rm Gr}}

\newbox\charbox
\newbox\slabox
\def\s#1{{      
        \setbox\charbox=\hbox{$#1$}
        \setbox\slabox=\hbox{$/$}
        \dimen\charbox=\ht\slabox
        \advance\dimen\charbox by -\dp\slabox
        \advance\dimen\charbox by -\ht\charbox
        \advance\dimen\charbox by \dp\charbox
        \divide\dimen\charbox by 2
        \raise-\dimen\charbox\hbox to \wd\charbox{\hss/\hss}
        \llap{$#1$}
}}

\def\spa#1.#2{\langle#1\,#2\rangle}
\def\spb#1.#2{[#1\,#2]}
\def\lor#1.#2{\left(#1\,#2\right)}

\def\spba#1.#2.#3{[ #1  | K_{#2} | #3 \ra  }
\def\spaa#1.#2.#3.#4{\la #1 | K_{#2} K_{#3} | #4 \ra }
\def\s#1{s_{#1}}
\def\t#1{t_{#1}}

\catcode`@=11  

\def\Slash#1{\hskip 0.05 cm \slash\hskip -0.22 cm #1}

\def\Tr{\, {\rm Tr}}

\def\eps{\epsilon}

\def\pol{\eps}

\def\la{\langle}
\def\ra{\rangle}

\def\lsl{\not{\hbox{\kern-2.3pt $\ell$}}}
\def\ksl{\not{\hbox{\kern-2.3pt $k$}}}

\def\rg{r_{\Gamma}}

\def\Aloop{A^{\rm 1-loop}}

\def\Atree{A^{\rm tree}}
\def\Aloop{A^{\rm 1-loop}}

\def\Li{\mathop{\hbox{\rm Li}}\nolimits}

\def\Li{\mathop{\hbox{\rm Li}}\nolimits}

\def\L{\left(}\def\R{\right)}

\def\dlips{d{\rm LIPS}}

\def\NeqEight{{\cal N} = 8}
\def\NeqFour{{\cal N} = 4}
\def\NeqOne{{\cal N} = 1}

\def\BRi#1#2#3{[ #1|{P}_{#3}|#2\ra}

\def\BRDM#1#2#3#4{\la#1| {#2} {#3}|#4 \ra}

\title{The Unitarity Method using a Canonical Basis Approach}

\author{David~C.~Dunbar, Warren~B.~Perkins and Edmund Warrick 
\\
Department of Physics,\\
Swansea University,\\
Swansea, SA2 8PP, UK }

\abstract{Various implementations of the Unitarity method have been developed to compute one-loop amplitudes in
gauge theories. In
this paper we present an implementation which uses canonical forms to generate
the rational coefficients of the basis integral functions.  
As an example, we present the results for the $\NeqOne$ contribution to
seven gluon scattering in closed, rational, analytic form.   }


\begin{document}
\section{Introduction}

One-loop computations are an essential ingredient in 
providing robust next-to-leading order predictions for QCD events at colliders such as the LHC~\cite{NLO}.  
A general one-loop amplitude for massless particles can be expressed, 
after an appropriate Passarino-Veltman reduction~\cite{Reduction}, in terms of $n$-point scalar integral functions $I_n^i$ with
rational coefficients, $a_i$, $b_j$, $c_k$, 
$$
 {\cal A}^{\rm 1-loop}_n =\sum_{i\in \cal C}\, a_i\, I_4^{i}
 +\sum_{j\in \cal D}\, b_{j}\, I_3^{j}
 +\sum_{k\in \cal E}\, c_{k} \,   I_2^{k}
+R_n +O(\eps) \,  .
\equn\label{DecompBasis}
$$ 
The functions $I_n^i$ contain all of the logarithms and dilogarithms in the amplitude  and $R_n$ is the remaining rational part.
The summations are over all possible integral functions. 
The Unitarity technique~\cite{BDDKa,BDDKb}   determines the rational coefficients of these functions from 
the information contained in the two-particle cuts shown in fig.\ref{cutfigure}, 
using physical on-shell amplitudes as inputs.

\noindent{The} cut, 
$$
C_{a,\ldots,b}  \equiv  
{ i \over 2} \int \dlips\biggl[ {\cal A}^{tree}(-\ell_1,a,a+1,\ldots,
b,\ell_2) \times {\cal A}^{tree}(-\ell_2,b+1,b+2,\ldots,a-1,\ell_1)
\biggr] \, ,
\equn
$$
\FIGURE[h]{
\resizebox*{5cm}{!}{\includegraphics{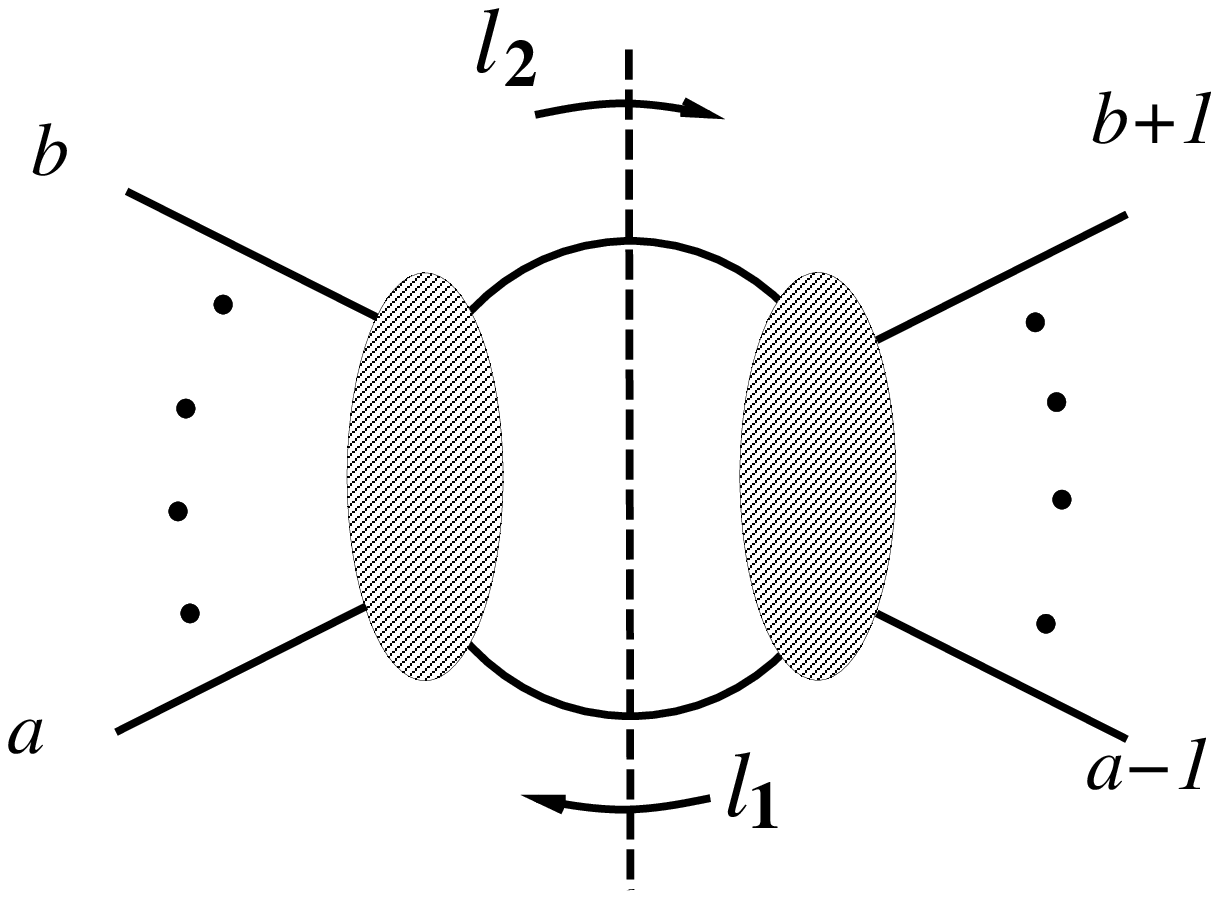}}
\caption{A two-particle cut of a one-loop amplitude}
\label{cutfigure}}

\noindent
where $\int \rm dLIPS$ denotes integration over the on-shell phase space of the $\ell_i$
is equal to the leading discontinuity in the integral
functions\cite{Cutkosky:1960sp} of eqn.~(\ref{DecompBasis}), 
$$
C_{a,\cdots,b}= \left(
\sum_{i\in \cal C'}\, a_i\, I_4^i +\sum_{j\in\,\cal D'}\, b_j\, I^j_3 +\sum_{k\in\,\cal E'}\, c_k\, I^k_2\,
\right) {\Big|}_{\rm Disc} \ ,
\equn\label{DecompDiscon}
$$
where ${\cal C'}$, ${\cal D'}$ and ${\cal E'}$ 
are the restricted sets of functions with a cut in this particular channel. In particular, ${\cal E'}$ consists of a single term.

The original implementation of the Unitarity method did not evaluate the cut directly but instead
manipulated the product of tree amplitudes, using $\ell_1^2=\ell_2^2=0$, to rewrite it in the form,
$$
\eqalign{
\int \dlips {\cal A}^{tree}(-\ell_1,a,a+1,\ldots,
b,\ell_2) \times {\cal A}^{tree}(-\ell_2,b+1,b+2,\ldots,a-1,\ell_1)
\cr
= \int \dlips \left( \sum  a_i { 1 \over (\ell_1-K_1^i)^2 (\ell_1-K_2^i)^2 } 
+\sum b_j { 1 \over (\ell_1-K_3^j)^2 }
+c_k \right) ,
\cr}\equn\label{OldUnitarity}
$$
and identified the coefficients in the above with the integral coefficients of eqn.(\ref{DecompDiscon}). A feature of the original implementation 
is that the representation~(\ref{OldUnitarity}) is not unique but one must simultaneously solve the full set of cut equations. 

In principle, the cut momenta $\ell_i$ 
should match the momenta of the integral functions $I_n$, i.e. they should be in $4-2\eps$ dimensions, however it was shown that for many amplitudes it is sufficient to use four dimensional tree amplitudes. Using four dimensional amplitudes allows us to evaluate the integral coefficients but not the rational terms $R_n$. For supersymmetric amplitudes, or the supersymmetric components of QCD amplitudes,  $R_n=0$ and we term these amplitudes ``cut-constructible''~\cite{BDDKa}.

In recent years considerable advances have been made in systematising the process of extracting the coefficients
of the basis integral functions.  Progress has been made both via the two-particle cuts above and using 
generalisations of unitarity~\cite{Bern:1997sc} where, for example, triple~\cite{Bidder:2005ri,Darren,BjerrumBohr:2007vu,Mastrolia:2006ki} and quadruple cuts~\cite{BrittoUnitarity} 
are utilised to identify the triangle and box coefficients.  Triple and quadruple cuts are useful in that they isolate contributions from smaller sets of integral functions. For example, a quadruple cut isolates a single box coefficient. Since the cut inserts four $\delta$-functions into the covariant integral the coefficient of this box function is given by the algebraic product of four tree amplitudes~\cite{BrittoUnitarity} . 
$\NeqFour$ one-loop amplitudes consist solely of scalar box functions and so quadruple cuts are sufficient to completely compute 
them~\cite{Bern:2004ky,Bern:2004jelly,Bern:2004llygdd,Bidder:2005in}:
a property shared by $\NeqEight$ supergravity amplitudes~\cite{NoTriangle}.

 The triple cut with three $\delta$-functions is effectively a one-parameter integral which can be evaluated via complex methods~\cite{Darren,BjerrumBohr:2007vu}. One may also consider one-particle cuts~\cite{NigelGlover:2008ur}.
Generalised Unitarity has been used beyond one-loop at two loops~\cite{Bern:1997nh,Bern:1998ug} and beyond~\cite{ThreeLoop}. 

Our strategy is to use all possible cuts and first evaluate the box coefficients from quadruple cuts then the triangle coefficients from triple cuts and finally use the two-particle cuts to determine the bubble coefficients. This is not the only strategy since the two-particle cuts contain enough information to determine the 
coefficients of the box and triangle contributions as well as the bubble coefficients. The testing ground for many of these techniques has been the computation of the various
terms in six gluon one-loop scattering amplitudes~\cite{Dunbar:2009uk} both for the supersymmetric 
contributions~\cite{BDDKa,BDDKb,Bidder:2004tx,BBCFsusyone} and also for
the ``cut-constructible'' parts of the QCD amplitudes~\cite{BDDKb,Bern:2005hh,Britto:2006sj}.  
For full QCD amplitudes the rational terms must also be calculated. Unitarity can be used to determine these, however this requires the use of tree amplitudes defined in $D=4-2\eps$ dimensions~\cite{Dunitarity}.
Alternatively, the rational pieces can be obtained using on-shell recursion~\cite{Berger:2006ci}, a method akin to that for tree amplitudes~\cite{Britto:2004ap}.
This has a numerical implementation together with Generalised Unitarity~\cite{blackhat}. Alternate numerical implementations exist for variants of this strategy~\cite{AlternateNumerical}. 
One may also use specialised Feynman diagram techniques which focus on the rational terms~\cite{Xiao:2006vt,Binoth:2006hk,Badger:2008cm}.

The approach we adopt recognises that there are a limited number of distinct structures that appear in the 
cut integrals. These may be evaluated in a number of ways: conventional covariant integration, fermionic integration~\cite{BBCFsusyone,Britto:2006sj},
direct extraction~\cite{Darren}
 or integrand level reduction~\cite{OPP}.  
By determining the contribution of each structure to the relevant coefficients we construct a 
canonical basis which, once constructed,  can be used for any amplitude.  
While this approach reproduces results from other methods, the decomposition into canonical forms is carried out directly on the 
four dimensional tree amplitudes 
without re-parametrising the $dLIPS$ integration. It produces compact, explicitly rational
expressions for the integral coefficients. 
We illustrate this process by presenting the $\NeqOne$ contribution to one-loop seven gluon scattering in closed analytic 
form\footnote{These are available in {\tt Mathematica} format at {\it http://pyweb.swan.ac.uk/$\sim$dunbar/sevengluon.html}}.

\section{Organisation of the Amplitudes}

The organisation of loop amplitudes into physical sub-amplitudes is an important step toward computing these amplitudes:
although eventually all the pieces must be reassembled. 
For one-loop amplitudes with adjoint particles, one may perform a {\it colour decomposition} similar to the tree-level
decomposition~\cite{ColorDecomposition}.  This one-loop decomposition
is~\cite{Colour},
$$
\eqalign{
{\cal A}_n^{\rm 1-loop}  = 
&i g^n \sum_{c=1}^{\lfloor{n/2}\rfloor+1}
      \sum_{\sigma \in S_n/S_{n;c}}
     \Gr_{n;c}\L \sigma \R\,A_{n;c}^{}(\sigma) \; ,
\cr}
\equn\label{ColourDecomposition}
$$
where ${\lfloor{x}\rfloor}$ is the largest integer less than or equal to $x$.\footnote{We have inserted a factor of $i$ in this definition to avoid universal factors of $i$ 
appearing in our
explicit formulae.}
The leading colour-structure factor,
$$
\Gr_{n;1}(1) \; = \; N_c\ \Tr\L T^{a_1}\cdots T^{a_n}\R \,,
\equn
$$
is just $N_c$ times the tree colour factor and the sub-leading colour
structures ($c>1)$ are given by,
$$
\Gr_{n;c}(1) \; = \; \Tr\L T^{a_1}\cdots T^{a_{c-1}}\R\,
\Tr\L T^{a_c}\cdots T^{a_n}\R \,.
\equn
$$
$S_n$ is the set of all permutations of $n$ objects
and $S_{n;c}$ is the subset leaving $\Gr_{n;c}$ invariant~\cite{Colour}.
The contributions from fundamental representation quarks circulating in the loop can be
obtained from the same partial amplitudes, except that the sum only runs 
over the $A_{n;1}$ and the overall factor of $N_c$ in $\Gr_{n;1}$ is
dropped.
For one-loop amplitudes of gluons the $A_{n;c}$, $c>1$ can be obtained from the $A_{n;1}$ by summing over 
permutations~\cite{Colour,BDDKa}. Hence it is sufficient to compute $A_{n;1}$ in what follows and, for clarity, we refer to these as $A_n$.  
The partial amplitudes $A_{n}$
have cyclic symmetry rather than full crossing symmetry.

We choose to use a {\it supersymmetric decomposition}. Instead of calculating the one-loop
contributions from massless gluons, $A_n^{[1]}$, or quarks, $A_n^{[1/2]}$, circulating in the loop, it is 
considerably more convenient to calculate the
contributions from the full $\NeqFour$ multiplet, a $\NeqOne$ chiral multiplet and a complex scalar circulating in the loop. In terms of these,
$$
\eqalign{
A_{n}^{[1]} &\; = \; A_{n}^{\,\NeqFour}-4A_{n}^{\,\NeqOne\; {\rm chiral}}\;+\;A_{n}^{[0]}\,,
\cr
A_{n}^{[1/2]} &\; = \; A_{n}^{\,\NeqOne\; {\rm chiral}}\;-\;A_{n}^{[0]}\,.
\cr}
\equn\label{SusyQCDDecomp}
$$

The amplitudes are also organised according to the helicities of the outgoing gluons which may be $\pm$. We
use polarisation tensors formed from Weyl spinors~\cite{Xu:1986xb},
$$
\pol^{+}_\mu (k;q) =
{ \la q^- |\gamma_\mu| k^- \ra
\over  \sqrt2 \spa{q}.k } \,, \hskip 0.1 cm  
\pol^{-}_\mu (k;q) =
{  \la q^+ |\gamma_\mu| k^+ \ra 
\over \sqrt{2} \spb{k}.q  } \,,
\equn
$$
where $k$ is the gluon momentum and $q$ is an arbitrary null
`reference momentum' which drops out of the final gauge-invariant
amplitudes.  The plus and minus labels on the polarization vectors
refer to the gluon helicities and we use the notation
$\langle ij \rangle\equiv  \langle k_i^{-} \vert k_j^{+} \rangle\, ,
[ij] \equiv \langle k_i^{+} \vert k_j^{-} \rangle$. In twistor-inspired studies of gauge theory amplitudes~\cite{Witten:2003nn}
the two component Weyl spinors are often expressed as,
$$
\lambda_a \equiv  |k^+\ra \; , \;\; \; \bar\lambda_{\dot a} \equiv |k^-\ra.
\equn
$$ 
Helicity amplitudes are related to those with all legs of opposite helicity by conjugation:
$\spa{a}.b \leftrightarrow \spb{b}.a$. 
Using spinor helicity
leads to amplitudes which are functions of the spinor variables $\spa{a}.b$ and
$\spb{a}.{b}$ and combinations such as $\la k_i^+| \Slash{p}|k_j^+\ra = [k_i|p|k_j\ra  \equiv\spb{k_i}.p\spa{p}.{k_j}$. It is useful to define combinations of spinor products,
$$
[a|P_{b\cdots f}|m\ra \equiv
\spb{a}.b\spa{b}.m +\cdots +
\spb{a}.f\spa{f}.m \, .
\equn
$$

In this article we  complete the computation of the $\NeqOne$ contributions to seven gluon scattering. 
Up to conjugation and relabeling,  there are nine independent helicity configurations for the colour ordered amplitudes.
The amplitudes $A_7(1^+,2^+,3^+,4^+,5^+,6^+,7^+)$ and $A_7(1^-,2^+,3^+,4^+,5^+,6^+,7^+)$ vanish to all orders 
in perturbation theory within any supersymmetry theory so,
$$
A^{\NeqOne}_7(1^+,2^+,3^+,4^+,5^+,6^+,7^+)=A^{\NeqOne}_7(1^-,2^+,3^+,4^+,5^+,6^+,7^+)=0 \; .
\equn
$$
Amplitudes with exactly two negative helicities are referred to as MHV (``maximally helicity violating'') amplitudes and those with three negative helicities as NMHV 
(``next to MHV'') amplitudes.  There are three  independent MHV and four independent NMHV helicity configurations for the seven gluon amplitude.  
The seven-point MHV amplitudes~\cite{BDDKb} and the NMHV amplitude with the three negative helicity legs adjacent  
$A^{\NeqOne}_7(1^-,2^-,3^-,4^+,5^+,6^+,7^+)$~\cite{Bidder:2005ri} are specific cases of known all-$n$ expressions. 
We present explicit forms for the remaining three NMHV partial amplitudes which will be made available 
at {\it http://pyweb.swan.ac.uk/$\sim$dunbar/sevengluon.html}
in  {\tt Mathematica} format.

\section{Canonical Basis for Bubble Coefficients from Two-Particle Cuts}

Consider the general decomposition of the product of tree amplitudes appearing in a two-particle cut
in terms of  canonical forms ${\cal F}_i$, 
$$ 
A^{\rm tree}(-\ell_1, \cdots, \ell_2) \times A^{\rm tree}(-\ell_2,\cdots, \ell_1)  = 
\sum c_i {\cal F}_i ({\ell_j}),
\equn
$$
where the $c_i$ are coefficients independent of $\ell_j$.  The forms ${\cal F}_i$ must have zero spinor weight in the $\ell_j$, i.e. 
they must be invariant under
$|\ell_j\ra \longrightarrow e^{i\phi_j}|\ell_j\ra, |\ell_j] \longrightarrow e^{-i\phi_j}|\ell_j]$. 
Each two-particle cut receives contributions from box functions, triangle functions and a bubble.  
The coefficients of the box functions are most simply determined from quadruple cuts and the triangle coefficients from 
triple cuts, so we only use the two-particle cut to determine the bubble coefficients.  
Consequently, we must determine the contributions to the 
bubble coefficients from the various canonical forms, ${\cal F}_i$.

We examine the ${\cal F}_i$ according to their leading order in the cut momenta $\ell$.
For a generic QCD amplitude the contributions 
have a maximum order of $\ell^2$.
In the $\NeqOne$ contributions to Yang-Mills amplitudes cancellations generically reduce this to order $\ell^0$.
Individual terms of order $\ell^N$ with $N < 0$ give no contribution to the bubble coefficients.  
To see this, we manipulate the cut into terms of the form,
$$
\int d^D \ell 
\delta(\ell^2)\delta((\ell-P)^2)
{
P^{N+R}(\ell)
\over \prod_{i=1}^R (\ell+Q_i)^2 },
\equn
$$
where the $Q_i$ may be null or non-null. For $N<0$,  Passerino-Veltman reduction on the corresponding covariant integral yields box and triangle integrals only.
Consequently we are only interested in terms of order $\ell^N$ with $N \geq 0$.

\subsection{Order $\ell^0$ Terms} 

These are the canonical forms which are required to obtain the
$\NeqOne$ contributions to Yang-Mills amplitudes.

Functionally we start with the simplest non-trivial case,
$$
\eqalign{
{\cal H} _1(A ; B ; \ell_1) &\equiv { \spa{\ell_1}.{B} \over \spa{\ell_1}.{A} }
=-{ [A|\ell_1|B\ra \over (\ell_1-k_A)^2 }, 
\cr}\equn
$$ 
where $k_A$ is taken to be real.
There are many ways to evaluate the contribution to the bubble coefficient from this simple form. We chose to
manipulate this as if it were a covariant integral. This means effectively
replacing, 
$$
\int dLIPS \longrightarrow  
 \int { d^D\ell  \over \ell_1^2 \ell_2^2 }\, , 
\equn
$$
then evaluating the covariant integral 
only keeping the coefficient of $-\ln(-P^2)$ in the  result.  
The integral  is then a linear triangle integral with massless leg  $k_A$ as shown, 
\begin{center}

\begin{picture}(80,60)(0,0)

\Line(30,15)(0,45)
\ArrowLine(60,45)(30,15)
\Line(0,45)(20,45)
\ArrowLine(20,45)(60,45)

\Line(30,15)(30,0)

\Text(-35,45)[]{$P-k_A$}
\Text(90,45)[]{$-P$}
\Text(21,0)[]{$k_A$}

\Line(60,45)(75,60)
\Line(60,45)(75,30)
\Line(60,45)(75,45)

\Line(0,45)(-15,60)
\Line(0,45)(-15,30)
\Line(0,45)(-15,45)
\CCirc(0,45){8}{Black}{Purple}
\CCirc(60,45){8}{Black}{Purple}

\SetWidth{2}
\DashCArc(79,55)(50,180,240){4}

\Text(52,27)[]{$\ell_1$}
\Text(37,53)[]{$\ell_2$}

\end{picture}
\end{center} 
which can be evaluated
by shifting the momenta using Feynman parameters,
$$
\ell_1^\mu \longrightarrow  \ell_1^\mu -k_A^\mu a_2 +P^\mu a_3,
\equn
$$
where $P\equiv k_a+k_{a+1}+\cdots k_b$ is the total momentum across the cut and $a_3$ refers to the Feynman parameter of the $\ell_2$ propagator. 

The $a_2$ term drops out of the integral and we use, 
$$
I_3[a_3]=
{ \ln( -P^2) -\ln( -(P-k_A)^2 ) \over 2k_A \cdot P }. 
\equn
$$
The coefficient of $-\ln(-P^2)$ is the bubble coefficient and thus ${\cal H}_1$ evaluates to $H_1$, 
$$
H_1[A ; B ; P ] = {[A|P|B \ra \over 2k_A\cdot P } ={[A|P|B \ra \over [A|P|A\ra}.
\equn
$$
Note we assume in the above that $k_{A}$ is real. If $A$ denoted a complex combination of momenta then $|A]$ should be replaced by $|A\ra^*$ in the canonical
form. 
We also have the conjugate result,
$$
{\bar{\cal  H}} _1 ( C ; D ; \ell_1 ) = { \spb{D}.{\ell_1} \over \spb{C}.{\ell_1} }
\Longrightarrow 
\bar H_1 [ C ; D  ; P ] =   {[D |P| C \ra \over [C|P|C\ra}.
\equn
$$
These forms satisfy,
$$
H_1[\; P|A] ; B ; P ] =  {1 \over P^2 }  \bar H_1 [ A ; P |B \ra  ; P ]. 
\equn
$$

We have chosen to determine the contribution from this form using covariant integrals, of course one obtains the same result by applying fermionic integration
~\cite{BBCFsusyone} or direct evaluation~\cite{Darren}. 
Our aim is to use this  simple result for $H_1$ as the starting point for generating many
further terms.   First consider ${\cal H}$ to be a holomorphic function of one of the $\ell_i$, i.e. ${\cal H}={\cal H}(|\ell\ra )$, then we can define,
$$
{\cal H}_n  ( A_i ; B_j ; \ell ) = { \prod_{j=1}^n \spa{B_j}.{\ell} \over \prod_{i=1}^n \spa{A_i}.\ell }, 
\;\;\;\; \spa{A_i}.{A_j} \neq 0 .
\equn
$$
By spinor weight the number of $A_i$ is the number of $B_j$. 
The cases where multiple poles appear are treated separately: such terms cancel in {\it amplitudes} by the factorisation theorems. 

By expressing this as a partial fraction
we can split ${\cal H}_n$ algebraically into a sum of ${\cal H}_1$ terms,
$$
{\cal H}_n( A_i ; B_i ; \ell )
=\sum_i c_i  { \spa{B_1}.{\ell}  \over \spa{A_i}.{\ell}  }
=\sum_i c_i  {\cal H}_1( A_i ; B_1 ;\ell ),
\equn
$$
where the coefficients $c_i$ are given by,
$$
c_i =  
{ \prod_{j=2}^n  \spa{B_j}.{A_i} \over \prod_{j\neq i} \spa{A_j}.{A_i} }. 
\equn
$$
The bubble coefficient generated by ${\cal H} _n$  is thus,
$$
H_n[ A_i ; B_i ; P ]= \sum_i { \prod_{j=2}^n  \spa{B_j}.{A_i} \over \prod_{j\neq i} \spa{A_j}.{A_i} } 
{\la B_1 | P | A_i ]   \over \la A_i | P | A_i ]  }.
\equn
$$

The same formula applies whether we have a holomorphic expression in $\ell_1$ or $\ell_2$.  
When we have a mixed expression, 
$$
{ \prod_{i=1}^n \spa{B_i}.{\ell_1} \over \prod_{i=1}^n \spa{A_i}.{\ell_1} }
{ \prod_{j=1}^m \spa{C_j}.{\ell_2} \over \prod_{j=1}^m \spa{D_j}.{\ell_2} } 
{ \prod_{k=1}^p \spb{E_k}.{\ell_1} \over \prod_{k=1}^p \spb{F_k}.{\ell_1} }
{ \prod_{l=1}^q \spb{G_l}.{\ell_2} \over \prod_{l=1}^q \spb{H_l}.{\ell_2} }, 
\equn
$$
we first rewrite it as a holomorphic  expression in $\ell_1$ and $\ell_2$,  
$$
{ \prod_{i=1}^n \spa{B_i}.{\ell_1} \over \prod_{i=1}^n \spa{A_i}.{\ell_1} }
{ \prod_{j=1}^m \spa{C_j}.{\ell_2} \over \prod_{j=1}^m \spa{D_j}.{\ell_2} } 
{ \prod_{k=1}^p [E_k|P|\ell_2\ra   \over \prod_{k=1}^p [F_k|P|\ell_2\ra   }
{ \prod_{l=1}^q [G_l|P|\ell_1\ra   \over \prod_{l=1}^q [H_l|P|\ell_1\ra   }, 
\equn
$$
then use the identity,
$$
{ \spa{a}.{\ell_2} \over  \spa{b}.{\ell_2} }
=
{ \spa{a}.{\ell_1}  \over  \spa{b}.{\ell_1}  }
-{ P^2\spa{a}.b  \over  \spa{b}.{\ell_1} [\ell_1|P | b \ra },
\equn
$$
to replace it by,
$$
{ \prod_{i=1}^n \spa{B_i}.{\ell_1} \over \prod_{i=1}^n \spa{A_i}.{\ell_1} }
{ \prod_{j=1}^m \spa{C_j}.{\ell_1} \over \prod_{j=1}^m \spa{D_j}.{\ell_1} } 
{ \prod_{k=1}^p [E_k|P|\ell_1\ra   \over \prod_{k=1}^p [F_k|P|\ell_1\ra   }
{ \prod_{l=1}^q [G_l|P|\ell_1\ra   \over \prod_{l=1}^q [H_l|P|\ell_1\ra   }
+\hbox{\rm terms of order $\ell^{-2}$ }.
\equn
$$
Only the leading term contributes to the bubble coefficient and it gives an overall contribution  
of $H_{n+m+p+q}[ A_i,D_j,P|F_k],P|H_l] ; B_i,C_j,P|E_k],P|G_l]; P ]$, 
provided the $A_i$, $D_j$, $P|F_k]$ and $P|H_l]$   are all distinct. 

We can have terms with $D=A$. In this case  we decompose into terms of the form,
$$
{\cal H}_2^x( A, A ; B_1 ,B_2  )={\spa{B_1}.{\ell_1}\spa{B_2}.{\ell_2}\over \spa{A}.{\ell_1}\spa{A}.{\ell_2} }.
\equn
$$
This special case canonical form gives, 
$$
{ H}_2^x[ A, A ; B_1 ,B_2 ; P ]= { [A|P|B_1\ra [A|P|B_2\ra\over [A|P|A\ra^2 }.
\equn
$$

There is a second class of functions arising from
terms with propagators involving non-null momenta such as, 
$$
{\cal G}_0 ( B ; D ; Q; \ell_1 ) = { 1 \over (\ell_1+Q)^2  } [ D | \ell_1 | B\ra,
\equn
$$
where $Q^2\neq 0$.
We can relate this to the ${\cal H}_1$ form using the  identity, 
$$
{ 1 \over (\ell+Q)^2  } \spb{D}.{\ell}
= { 1 \over (\ell+Q)^2  } { [D|P(P+Q)Q|\ell\ra \over \la \ell | PQ | \ell \ra }
-{ [D|P|\ell\ra \over \la \ell | PQ | \ell \ra } \, , 
\equn
$$
leading to,
$$
{ 1 \over (\ell_1+Q)^2  } [ D | \ell_1 | B\ra
=
-{ [D|P|\ell_1\ra \spa{B}.{\ell_1}
\over \la \ell_1 | PQ | \ell_1 \ra } 
+\hbox{\rm sub-leading}. 
\equn\label{G0splitEQ}
$$
We then make the replacement,
$$
\la \ell | PQ | \ell \ra 
\sim\la \ell | \hat P\hat Q | \ell \ra 
=\spa{\ell}.{\hat P} \spb{\hat P}.{\hat Q}\spa{\hat Q}.\ell,
\equn\label{PQdefnEQ}
$$
where $\hat P$ and $\hat Q$ are the null linear combinations of $P$ and $Q$, 
$$
\hat P^\mu={1\over 2\sqrt{\Delta_3}}\bigl( P^2 Q^\mu -(P\cdot Q-{ \sqrt{\Delta_3}\over 2} )P^\mu\bigr),
\quad
\hat Q^\mu={1\over 2\sqrt{\Delta_3}}\bigl(-P^2 Q^\mu +(P\cdot Q+{ \sqrt{\Delta_3}\over 2})P^\mu\bigr),
\equn\label{PQdefnBEQ}
$$
with $\Delta_3=\Delta_3(P,Q)=4(P\cdot Q)^2-4P^2Q^2$. $\Delta_3$ is the Gram determinant of the three-mass triangle integral having legs of momenta $P$, $Q$ and $-P-Q$. 
The leading term in eqn.~(\ref{G0splitEQ})
then has precisely the form of an ${\cal H}_2$ function. 
Splitting this into a pair of ${\cal H}_1$ functions gives two terms 
that are not individually  rational because $\hat P$ and $\hat Q$ contain
factors of $\sqrt{\Delta_3}$. The pair of terms are however the irrational conjugates of each other,
so the sum is rational. To have canonical forms which yield explicitly rational coefficients  we choose to evaluate ${\cal G}_0$ as a separate canonical form 
which is manifestly rational:
$$
G_0 [ B ; D ; Q; P ] 
={  [D |  P(QP-PQ) | B \ra \over \Delta_3 }=
{  [D |  P[Q,P] | B \ra \over  \Delta_3 }. 
\equn
$$

We commonly find the form, 
$$
{\cal G}_1( A ; B_0, B_1 ; D ; Q ; \ell_1 )= { 1 \over (\ell_1+Q)^2  } 
{ [ D | \ell_1 | B_0\ra \spa{\ell_1}.{B_1} \over \spa{\ell_1}.{A} }.
\equn 
$$
For $\spa{A}.{\hat P}, \spa{A}.{\hat Q}\neq 0$, this can be decomposed as 
an ${\cal H}_3$ function but at the cost of introducing 
irrational coefficients. Once again, combining these terms generates a manifestly 
rational form. For $\spa{A}.{\hat P}, \spa{A}.{\hat Q}\neq 0$,
$$
\eqalign{{ G}_1[ A ; B_0, B_1 ; D ; Q ; P ]
& =  -{[D|P [P,Q]| A\ra\la B_1|[P,Q]| B_0\ra
  \over 2\Delta_3\la A|PQ| A\ra}
\cr 
&  + 
{[D|P|A\ra  ( \spa{B_0}.{A}\la B_1|P|A]+\spa{B_1}.{A}\la B_0|P|A] )  \over 2\la A|PQ|A\ra \la A|P|A]} \, .
\cr}
\equn
$$ 
We can extend this form to determine the bubble contributions arising from,  
$$
{\cal G}_n( A_i ; B_0, B_i ; D ; Q ; \ell_1 )= { 1 \over (\ell_1+Q)^2  } 
{ [ D | \ell_1 | B_0\ra \prod_{i=1}^n\spa{\ell_1}.{B_i} \over\prod_{i=1}^n \spa{\ell_1}.{A_i} },
\equn
$$
by splitting it into a sum of ${\cal G}_1$ terms, just as we split ${\cal H}_n$ 
into a sum of ${\cal H}_1$ terms, 
$$
{\cal G} _n( A_i ; B_0, B_i ; D ; Q ; \ell_1 )
=\sum_i  c_i\, {\cal G}_1( A_i ; B_0, B_n ; D ; Q ; \ell_1 ), 
\equn
$$
with,
$$
c_i =  
{ \prod_{j < n }  \spa{A_i}.{B_j} \over \prod_{j\neq i} \spa{A_i}.{A_j} }. 
\equn
$$

The $H_n$ and $G_n$ functions are sufficient to evaluate the bubble
coefficients of the $\NeqOne$ contributions to Yang-Mills amplitudes.
We will show this by example in
the following sections where we explicitly evaluate the seven-point
$\NeqOne$ one-loop contributions. 

In general we may also have terms with multiple  propagators involving non-null momenta,
$$
{ f(\ell)  \over (\ell+Q_1)^2 (\ell+Q_2)^2 \ldots}.
\equn
$$
Using the constraint $(\ell-P)^2=0$, we have,
$$
(\ell+Q_1)^2=[\ell|Q_1+{Q_1^2\over P^2}P|\ell\ra \equiv [\ell|{\cal Q}_1|\ell\ra,
\equn
$$
where ${\cal Q}_1$ is a non-null linear combination of $Q_1$ and $P$. 
This
allows any multiple propagator terms to be written as,
$$
{ f(\ell)  \over [\ell|{\cal Q}_1|\ell\ra [\ell|{\cal Q}_2|\ell\ra \ldots}.
\equn
$$
Partial fractioning on the $\bar\lambda(\ell)$'s then gives a sum of terms of the form,
$$
{g(\ell)\over[\ell|{\cal Q}_1|\ell\ra \la \ell|{\cal Q}_1{\cal Q}_2|\ell\ra \ldots}
\sim
{g(\ell)\over[\ell|{\cal Q}_1|\ell\ra \la     \ell          \hat{\cal Q}_1 \ra 
                              [  \hat{\cal Q}_1 \hat{\cal Q}_2  ] 
                             \la \hat{\cal Q}_2     \ell          \ra \ldots},
\equn
$$
which  can be further split into a sum of ${\cal G}_1$ forms at the expense of introducing irrational factors in 
$\sqrt{\Delta_3({\cal Q}_i,{\cal Q}_j)}$. As with the $\sqrt{\Delta_3(P,Q)}$ terms, 
these all arise in irrational conjugate pairs and
the sum is  rational.

\subsection{Terms of Order $\ell^1$ and $\ell^2$} 

For the scalar contributions to Yang-Mills amplitudes we need forms of order $\ell^1$ and $\ell^2$. 
In general we will needs forms denoted $H_n^r$ for contributions
of order $\ell^r$ where
the denominator has $n$ factors of $\spa{A_i}.\ell$. 

The higher order ${\cal H}$ and $H$ forms are:

$$
\eqalign{
{\cal H}_0^1 &\equiv [ D|\ell_1|B\ra
\to  H_0^1[B; D; P] = {1 \over 2} [D|P|B\ra, 
\cr
{\cal H}_1^1 &\equiv { [D|\ell_1|B_1\ra \spa{\ell_1}.{B_2} \over \spa{\ell_1}.A }
\to  
\cr
& 
H_1^1[ A; B_1,B_2; D ; P]  = { P^2 \over 4  [A|P|A\ra^2 }\left( 
[D|A|B_1\ra[A|P|B_2\ra+ [D|A|B_2\ra[A|P|B_1\ra \right)
\cr
&\hskip 60pt
+{ 1 \over 4  [A|P|A\ra }
\left([D|P|B_1\ra [A|P|B_2\ra+[D|P|B_2\ra [A|P|B_1\ra \right),
\cr
{\cal H}_0^2 &\equiv [D_1|\ell_1|B_1\ra [D_2|\ell_1|B_2\ra \to 
\cr
& H_0^2[ B_1,B_2; D_1,D_2 ; P] ={1 \over 3} [D_1|P|B_1\ra[D_2|P|B_2\ra
+{P^2 \over 6}  \spb{D_1}.{D_2}\spa{B_1}.{B_2}  ,
\cr
{\cal H}_1^2 & \equiv  { [D_1|\ell_1|B_1\ra [D_2|\ell_1|B_2\ra \spa{\ell_1}.{B_3} \over \spa{\ell_1}.{A} } 
\to 
\cr
&  H_1^2[ A; B_1,B_2,B_3; D_1,D_2  ; P] = 
{ (P^2)^2 \over 18 [A|P|A\ra^3 } \left ( 
[D_1|A|B_1\ra [D_2|A|B_2\ra[A |P |B_3\ra+ {\cal P}_6(B_i) 
\right)
\cr
&\hskip 1.0 truecm +
{ (P^2) \over 36 [A|P|A\ra^2 } \left( 
[D_1|P|B_1\ra [D_2|A|B_2\ra[A |P |B_3\ra
+{\cal P}_{12} ( B_i,D_i)\right )
\cr
&\hskip 1.0 truecm +
{ 1 \over 18 [A|P|A\ra } \left( 
[D_1|P|B_1\ra[D_2|P|B_2\ra[A |P |B_3\ra+ {\cal P}_6(B_i) \right)\, ,
\cr}
\equn
$$
where ${\cal P}_6(B_i)$ and ${\cal P}_{12} ( B_i,D_i)$ represent a total of six and twelve 
permutations respectively.

\def\zA{D}
\def\zB{C}
\def\zC{B}
\def\zD{A}

The higher order terms involving  propagators with non-null momenta can be reduced to ${\cal G}_1^2$ or ${\cal G}_1^1$
forms, where,
$$
{\cal G}^n_1 \equiv  f^n(\ell)
{\spb{\zA}.{\ell}\spa{\zB}.{\ell}\spa{\zC}.{\ell}\over [\ell|{\cal Q}|\ell\ra \spa{\zD}.{\ell}},
\equn
$$
and $f^n(\ell)$ is a polynomial of degree $n$ in $\ell$.
To evaluate these forms we again make use of the identity:
$$
{\spb{D}.{\ell}\over [\ell|{\cal Q}|\ell\ra}
=
P^2{[D|{\cal Q}|\ell\ra\over [\ell|{\cal Q}|\ell\ra\la \ell|P{\cal Q}|\ell\ra} -{[D|P|\ell\ra\over\la \ell|P{\cal Q}|\ell\ra}.
\equn
$$
This allows us to write,
$$
\eqalign{
{\cal G}^n_1 &\la \zD|P{\cal Q}|\zD\ra
\cr
= &
{P^2f(\ell)\spa{\zC}.{\ell}\over [\ell|{\cal Q}|\ell\ra}\left( 
 {[\zA|{\cal Q}|\zD\ra\spa{\zB}.{\zD}\over\spa{\zD}.{\ell}}
+{4\over P^2}
\left(
 {[\zA|{\cal Q}|\hat P\ra \la \zB|\hat P\hat {\cal Q}|\zD\ra\over \la \ell \hat P\ra}
+{[\zA|{\cal Q}|\hat {\cal Q}\ra \la \zB|\hat {\cal Q}\hat P|\zD\ra\over \la \ell \hat {\cal Q}\ra}
\right)\right)
\cr &
-f(\ell)\spa{\zC}.{\ell}\left(
{[\zA|P|\zD\ra\spa{\zB}.{\zD}\over\spa{\zD}.{\ell}}
+{4\over P^2}
\left(
 {[\zA|P|\hat P\ra\la \zB|\hat P \hat {\cal Q}|\zD\ra\over\la \ell\hat P\ra}
+{[\zA|P|\hat {\cal Q}\ra\la \zB|\hat {\cal Q} \hat P|\zD\ra\over\la \ell\hat {\cal Q}\ra}
\right)\right) .
\cr}
\equn
$$
This splits the term of interest into pieces with the same overall power 
count in $\ell$ but only massless propagators and terms with a reduced
power count in $\ell$ involving the original  propagator. The former 
are readily evaluated using our basis of $H$ functions, while the latter
rely on $G$ functions. The terms involving the propagator along
with $\spa{\hat P}.{\ell}^{-1}$ or $\spa{\hat {\cal Q}}.{\ell}^{-1}$ give rise to special 
cases of the $G$ functions as discussed below.
Using this splitting procedure, the order $\ell^1$ term, ${\cal G}_1^1$,
is expressed in terms of a sum involving special cases of the order $\ell^0$  ${\cal G}_1$ form,
namely, ${\cal G}_1^{\hat Q}$ and ${\cal G}_1^{\hat P}$, where,
$$
{\cal G}_1^{\hat Q}={\spb{\zA}.{\ell}\spa{\zB}.{\ell}\spa{\zC}.{\ell}\over [\ell|{\cal Q}|\ell\ra \spa{\hat {\cal Q}}.{\ell}} \; .
\equn
$$
Evaluating this form gives the bubble contribution,
$$
G_1^{\hat Q}= 
2 {
[\zA|[{\cal Q},P]P|\zB\ra[\hat {\cal Q}|P|\zC\ra
+\bigl(P^2[\zA|{\cal Q}|\zC\ra-(P.{\cal Q}+\sqrt{\Delta_3}/2 )[\zA|P|\zC\ra\bigr)[\hat {\cal Q}|P|\zB\ra
\over
P^2\Delta_3}.
\equn
$$
 ${\cal G}_1^{\hat P}$ and  $G_1^{\hat P}$ are obtained by irrational conjugation.

Similarly for the order $\ell^2$ terms we require special cases of ${\cal G}_1^1$  with  $\spa{D}.{\hat Q}=0$ 
and  $\spa{D}.{\hat P}=0$. 
Setting $f(\ell)=[E|\ell|F\ra$, we have,
$$
{\cal G}_1^{1:\hat Q}=
{[E|\ell|F\ra [\zA|\ell|\zB\ra \spa{\ell}.{\zC}
  \over [\ell|{\cal Q}|\ell\ra\spa{\ell}.{\hat {\cal Q}}}.
\equn
$$
Making the definitions,
$$
\eqalign{
\beta = \biggl(\qquad & [E|\hat P|F\ra [\zA|\hat P|\zB\ra [\hat {\cal Q}|P|\zC\ra
                        -[E|\hat {\cal Q}|F\ra [\zA|\hat {\cal Q}|\zB\ra [\hat {\cal Q}|P|\zC\ra 
\cr -&
               [E|\hat {\cal Q}|\zC\ra [\zA|\hat {\cal Q}|\zB\ra [\hat {\cal Q}|P|F\ra
               -[\zA|\hat {\cal Q}|\zC\ra [E|\hat {\cal Q}|F\ra [\hat {\cal Q}|P|\zB\ra
\biggr),
\cr}
\equn
$$
and
$$
\eqalign{
\gamma=\biggl(
   \Bigr(& [E|\hat P|F\ra-  [E|\hat {\cal Q}|F\ra\Bigr)
   \Bigr(  [\zA|\hat P|\zB\ra-  [\zA|\hat {\cal Q}|\zB\ra\Bigr)[\hat {\cal Q}|P|\zC\ra
\cr &
  -\Bigl( [E|\hat P|\zB\ra[\zA|\hat {\cal Q}|F\ra
         +[E|\hat {\cal Q}|\zB\ra[\zA|\hat P|F\ra\Bigr)[\hat {\cal Q}|P|\zC\ra
\cr &
  -[E|\hat {\cal Q}|\zC\ra\Bigr([\zA|\hat P|\zB\ra-[\zA|\hat Q|\zB\ra\Bigr)[\hat Q|P|F\ra
\cr &
  -[\zA|\hat Q|\zC\ra\Bigr([E|\hat P|F\ra-[E|\hat Q|F\ra\Bigr)[\hat Q|P|\zB\ra
\biggr),
\cr}
\equn
$$
we find that this form gives a contribution to the bubble coefficient of,
$$
G_1^{1:\hat Q}=
 -8{\sqrt{\Delta_3}\beta+P.{\cal Q}\gamma \over P^2\Delta_3}.
\equn
$$
Again ${\cal G}_1^{1:\hat P}$ and  $G_1^{1:\hat P}$ are obtained by irrational conjugation.
 
Although these forms contain $\sqrt{\Delta_3}$, we are always interested in the sum of irrational-conjugate
pairs and the final canonical form is guaranteed to be rational.

\section{Example: Seven-point $\NeqOne$ Contributions}
\label{SevenPointSection}

The basic NMHV amplitudes are:
$$
\eqalign{
A: & \,A_7(1^-,2^-,3^+,4^-,5^+,6^+,7^+) 
\cr
B: & \,A_7(1^-,2^-,3^+,4^+,5^-,6^+,7^+) 
\cr
C: & \,A_7(1^-,2^+,3^-,4^+,5^-,6^+,7^+) 
\cr
D: & \,A_7(1^-,2^-,3^-,4^+,5^+,6^+,7^+) 
\cr}
\equn
$$
Amplitude $A$ has no permutation symmetries while amplitudes $B$ and $C$  have the following 
invariances:
$$
\eqalign{
B:  (1234567) \leftrightarrow (2176543) \cr
C:  (1234567) \leftrightarrow (5432176) \cr}
$$
Amplitude $D$ has the simplest structure and is an example of a ``split-helicity'' amplitude 
whose $n$-point expression is given in~\cite{Bern:2005hh}.

The supersymmetric cancellations present in a $\NeqOne$ computation lead to the one-loop amplitude being of the form,
$$
 A_n^{\NeqOne}=\sum_{i\in \cal C}\, a_i\, I_4^{i}
 +\sum_{j\in \cal D}\, b_{j}\, I_3^{j}
 +\sum_{k\in \cal E}\, c_{k} \,   I_2^{k}
 \; , 
\equn
$$ 
with no further rational terms.  Since all the coefficients can be evaluated from four-dimensional unitarity these contributions are termed ``cut constructible''.

For the contribution from the $\NeqOne$ chiral multiplet the coefficients of the box integral functions contain sufficient information to determine the coefficients of 
the one and two mass triangle functions. As discussed in the Appendix~\ref{IntegralFunctionsAp}, we choose to absorb these triangles into the box functions, 
leaving a basis
of {\it truncated} boxes ${\cal F}_4^i$, three mass triangles and bubble functions, 
$$
 A_n^{\NeqOne\; {\rm chiral}}=\sum_{i\in \cal C}\, a_i\, {\cal F}_4^{i\;}
 +\sum_{{j\in \cal D}_{3m}}\, b_{j}\, I_3^{3m \; j}
 +\sum_{k\in \cal E}\, c_{k} \,   I_2^{k}
\; .
\equn
$$

\subsection{ $A^{\NeqOne}_7(1^-,2^-,3^+,4^-,5^+,6^+,7^+)$}

\def\Ibub{I_2}

This amplitude can be decomposed into 20 integral functions:  
$$
\eqalign{
A^{\NeqOne\ \rm chiral}_7(1^-,2^-, &  3^+,4^-,5^+,6^+,7^+) =
a^A_1 {\cal F}^{3m }_{6\{71\}\{23\}\{45\}} 
\cr
&+a^A_2 {\cal F}^{2m\; h}_{1\{23\}\{456\}7}
+a^A_3 {\cal F}^{2m\; h}_{3\{45\}\{671\}2}
+a^A_4 {\cal F}^{2m\; h}_{3\{456\}\{71\}2}
+a^A_5 {\cal F}^{2m\; h}_{5\{671\}\{23\}4}
\cr
&+a^A_6 {\cal F}^{2m\; e}_{3\{45\}6\{712\}}
+a^A_7 {\cal F}^{1m}_{234\{5671\}}
+a^A_8 {\cal F}^{1m}_{345\{6712\}}
\cr
&+b^A_1 I^{3m\; tri}_{\{23\}\{45\}\{671\}}
+b^A_2 I^{3m\; tri}_{\{71\}\{23\}\{456\}}
\cr
&
+c^A_1 \Ibub (t_{123})+c^A_2 \Ibub (t_{234})
+c^A_3 \Ibub (t_{345})+c^A_4 \Ibub (t_{456})
+c^A_6 \Ibub (t_{671})
\cr
&+c^A_7\Ibub( t_{712})
\cr
&+d^A_2 \Ibub (s_{23} )+d^A_3 \Ibub (s_{34} )
+d^A_4 \Ibub (s_{45} )+d^A_7 \Ibub (s_{71} ) ,
\cr}
\equn
$$
where we have chosen to label the boxes and three-mass triangles by the clustering of the legs. 

\subsection{ $A^{\NeqOne}_7(1^-,2^-,3^+,4^+,5^-,6^+,7^+)$}

This amplitude can be decomposed into 25 integral functions:
$$
\eqalign{
A^{\NeqOne\ \rm chiral}_7(1^-,2^-, &  3^+,4^+,5^-,6^+,7^+) =
a^B_1 {\cal F}^{3m }_{6\{71\}\{23\}\{45\}} +a^B_2 {\cal F}^{3m }_{4\{56\}\{71\}\{23\}} 
\cr
&
+a^B_3 {\cal F}^{2m\; h}_{1\{23\}\{456\}7}
+a^B_4 {\cal F}^{2m\; h}_{3\{45\}\{671\}2}
+a^B_5 {\cal F}^{2m\; h}_{6\{71\}\{234\}5}
+a^B_6 {\cal F}^{2m\; h}_{1\{234\}\{56\}7}
\cr
&
+a^B_7 {\cal F}^{2m\; h}_{5\{671\}\{23\}4}
+a^B_8 {\cal F}^{2m\; h}_{3\{456\}\{71\}2}
\cr
&+a^B_9 {\cal F}^{2m\; e}_{3\{45\}6\{712\}}
 +a^B_{10} {\cal F}^{2m\; e}_{4\{56\}7\{123\}}
+a^B_{11} {\cal F}^{1m}_{456\{7123\}}
\cr
&+b^B_1 I^{3m\; tri}_{\{23\}\{45\}\{671\}}
+b^B_2 I^{3m\; tri}_{\{71\}\{23\}\{456\}}
+b^B_3 I^{3m\; tri}_{\{56\}\{71\}\{234\}}
\cr
&
+c^B_1 \Ibub (t_{123})+c^B_2 \Ibub (t_{234})
+c^B_3 \Ibub (t_{345})+c^B_4 \Ibub (t_{456})
+c^B_5 \Ibub (t_{567})
\cr
&+c^B_6 \Ibub (t_{671})+c^B_7\Ibub( t_{712})
\cr
&+d^B_2 \Ibub (s_{23} )+d^B_4 \Ibub (s_{45} )
+d^B_5 \Ibub (s_{56} )+d^B_7 \Ibub (s_{71} ).
\cr}
\equn
$$

\subsection{ $A^{\NeqOne}_7(1^-,2^+,3^-,4^+,5^-,6^+,7^+)$}

This amplitude can be decomposed into 37 integral functions:
$$
\eqalign{
 &A^{\NeqOne\ \rm chiral}_7(1^-,2^+, 3^-,4^+,5^-,6^+,7^+) =
a^C_1 {\cal F}^{3m }_{6\{71\}\{23\}\{45\}} +a^C_2 {\cal F}^{3m }_{7\{12\}\{34\}\{56\}} 
\cr
&+a^C_3 {\cal F}^{2m\; h}_{1\{23\}\{456\}7}
+a^C_4 {\cal F}^{2m\; h}_{2\{34\}\{567\}1}
+a^C_5 {\cal F}^{2m\; h}_{3\{45\}\{671\}2}
+a^C_6 {\cal F}^{2m\; h}_{4\{56\}\{712\}3}
+a^C_7 {\cal F}^{2m\; h}_{6\{71\}\{234\}5}
\cr
&
+a^C_8 {\cal F}^{2m\; h}_{1\{234\}\{56\}7}
+a^C_9 {\cal F}^{2m\; h}_{3\{456\}\{71\}2}
+a^C_{10} {\cal F}^{2m\; h}_{4\{567\}\{12\}3}
+a^C_{11} {\cal F}^{2m\; h}_{5\{671\}\{23\}4}
+a^C_{12} {\cal F}^{2m\; h}_{6\{712\}\{34\}5}
\cr
&
+a^C_{13} {\cal F}^{2m\; e}_{4\{56\}7\{123\}}
+a^C_{14} {\cal F}^{2m\; e}_{6\{71\}2\{345\}}
\cr
&
+a^C_{15} {\cal F}^{1m}_{123\{4567\}}
+a^C_{16} {\cal F}^{1m}_{234\{5671\}}
+a^C_{17} {\cal F}^{1m}_{345\{6712\}}
+a^C_{18} {\cal F}^{1m}_{456\{7123\}}
+a^C_{19} {\cal F}^{1m}_{712\{3456\}}
\cr
&+b^C_1 I^{3m\; tri}_{\{12\}\{34\}\{567\}}
+b^C_2 I^{3m\; tri}_{\{712\}\{34\}\{56\}}
+b^C_3 I^{3m\; tri}_{\{71\}\{23\}\{456\}}
+b^C_4 I^{3m\; tri}_{\{71\}\{234\}\{56\}}
+b^C_5 I^{3m\; tri}_{\{671\}\{23\}\{45\}}
\cr
&
+c^C_1 \Ibub (t_{123})+c^C_2 \Ibub (t_{234})
+c^C_3 \Ibub (t_{345})+c^C_4 \Ibub (t_{456})
+c^C_5 \Ibub (t_{567})
\cr
&+c^C_6 \Ibub (t_{671})+c^C_7\Ibub( t_{712})
\cr
&+d^C_1 \Ibub (s_{12} )+d^C_2 \Ibub (s_{23} )+d^C_3 \Ibub (s_{34} )+d^C_4 \Ibub (s_{45} )
+d^C_5 \Ibub (s_{56} )+d^C_7 \Ibub (s_{71} ).
\cr}
\equn
$$

Altogether these  NMHV $\NeqOne$ contributions are specified by 82 coefficients. 
The box and three-mass triangle coefficients are special
cases of generic forms which are given in appendix~\ref{BoxAppendix} and section~\ref{gen3masstrisection} respectively.  The twenty $c_i$
coefficients are either special cases of the generic $C_0$ function~(\ref{N1MHVMHVbar}) or are given by one of four forms, $C_X$, specific to
the seven-point case.   The remaining parts of the amplitudes are given by the fourteen
$d_i^\alpha$ functions, which are not all independent 
but can be expressed in terms of six $D_X$ functions.

\subsection{$C_X$ Functions}

We illustrate the calculation of the $C_X$ functions by considering an explicit realisation of the $C_B$ 
function which  arises as the coefficient of $\Ibub(t_{712})$ and is obtained by computing  the {$t_{712}$ cut  of $A_7(1,2,3^-,4^+,5^-,6^+,7)$},
where precisely one of legs $7$, $1$ or $2$ has negative helicity. We label this leg $m_1$.

The product of tree amplitudes generated by the cut is,
$$
\sum_{h} A^{\rm tree} (-\ell_1^{h}, 7,1,2,\ell_2^{-h}) \times A^{\rm tree}(-\ell_2^{h},3^-,4^+,5^-,6^+,\ell_1^{-h
}),
\equn\label{multipletsumEQ}
$$
where the summation is over the complex scalar and fermionic states of the $\NeqOne$ chiral multiplet. 
The six point NMHV tree amplitude has three terms, 
$$
 A^{\rm tree}(-\ell_2^{h},3^-,4^+,5^-,6^+,\ell_1^{-h})
 =T_{1}^h+T_{2}^h+T_{3}^h \; ,
\equn
$$ 
where,
$$
\eqalign{
T_{1}^h=&
{ \BRi{\ell_2}5{456}^2 \BRi{\ell_1}5{456}^2 \over
t_{456} \spb{\ell_1}.{\ell_2}\spb{\ell_2}.3\spa4.5\spa5.6 \BRi{\ell_1}4{456}\BRi36{456} }
\left( -{ \BRi{\ell_2}5{456} \over \BRi{\ell_1}5{456}} \right)^{h},
\cr
T_{2}^h=&
{ \BRi63{56\ell_1}^2(\spa{\ell_2}.3\spb6.{\ell_1})^2\over
t_{56\ell_1} \spa{\ell_2}.3\spa3.4\spb5.6\spb6.{\ell_1} \BRi5{\ell_2}{56\ell_1}\BRi{\ell_1}4{56\ell_1} }
 \left( -{ \BRi63{56\ell_1} \over \spa{\ell_2}.3\spb6.{\ell_1}} \right)^{h},
\cr
T_{3}^h=& 
{ \BRi4{\ell_1}{345}^2\BRi4{\ell_2}{345}^2
\over
t_{345} \spa6.{\ell_1}\spa{\ell_1}.{\ell_2}\spb3.4\spb4.5 \BRi36{345}\BRi5{\ell_2}{345} }
\left({ -\BRi4{\ell_1}{345}\over\BRi4{\ell_2}{345} }\right)^h, 
\cr}
\equn
$$
with $h$ denoting the helicity of the leg $\ell_2$: $h=0$ for a scalar and $h=\pm 1$ for a fermion. 

Summing over the multiplet in eqn.~(\ref{multipletsumEQ})  leads to term by term cancellations which we can express as $\rho$ factors multiplying 
each product of $h=0$ terms, leading to a cut integrand of the form,
$$
{ \spa{m_1}.{\ell_1}^2 \spa{m_1}.{\ell_2}^2 \over
\spa7.1\spa1.2 \spa2.{\ell_2}\spa{\ell_2}.{\ell_1} \spa{\ell_1}.7
}
\times 
\biggl( 
T_{1}^0  \,\rho_{1} +
T_{2}^0  \,  \rho_{2} +
T_{3}^0  \,  \rho_{3} 
\biggr),
\equn
$$
where 
the $\rho$ factors are,
$$
\eqalign{
\rho_{1}&
={\la m_1 | P_{712}P_{456}|5\ra^2
\over\spa{m_1}.{\ell_2}\BRi{\ell_2}5{456}\spa{m_1}.{\ell_1}\BRi{\ell_1}5{456}},
\cr
\rho_{2}&
= { \spa{Y_{B2}}.{\ell_2}^2
\over \spa{m_1}.{\ell_2}\BRi63{56\ell_1}  \spa{m_1}.{\ell_1} \spa{\ell_2}.3\spb6.{\ell_1}  },
\cr
\rho_{3}&
={  [ 4 | P_{345} |m_1\ra^2 \spa{\ell_1}.{\ell_2}^2  
\over  \spa{m_1}.{\ell_1}\BRi4{\ell_1}{345}\spa{m_1}.{\ell_2} \BRi4{\ell_2}{345}    },
\cr}
\equn
$$
with,
$$
|Y_{B2}\ra = \spb6.5\spa5.3 |m_1\ra +
\spb6.7\spa{m_1}.3 |7\ra+
\spb6.1\spa{m_1}.3 |1\ra+
\spb6.2\spa{m_1}.3 |2\ra.
\equn
$$
Consequently the contribution of the $T_{1}$ terms to the cut integrand is, 
$$
\eqalign{
& {  \spa{m_1}.{\ell_1}\spa{m_1}.{\ell_2}
\over \spa7.1 \spa1.2 \spa2.{\ell_2}\spa{\ell_2}.{\ell_1}\spa{\ell_1}.7 }
\times
{ \BRi{\ell_2}5{456}\BRi{\ell_1}5{456} \la m_1|P_{712}P_{456}|5\ra^2 \over
t_{456} \spb{\ell_1}.{\ell_2}\spb{\ell_2}.3\spa4.5\spa5.6 \BRi{\ell_1}4{456}\BRi36{456} }
\cr
& =
{\la m_1|P_{712}P_{456}|5\ra^2
\over 
 \spa7.1 \spa1.2\spa4.5\spa5.6 \BRi36{456}t_{456}t_{712}  }
\times
 {  \spa{m_1}.{\ell_1}\spa{m_1}.{\ell_2}
\over \spa2.{\ell_2}\spa{\ell_1}.7 }
\times
{ \BRi{\ell_2}5{456}\BRi{\ell_1}5{456}  \over
 \spb{\ell_2}.3 \BRi{\ell_1}4{456} }
\cr
& =
{\la m_1|P_{712}P_{456}|5\ra^2
\over 
 \spa7.1 \spa1.2\spa4.5\spa5.6 \BRi36{456}t_{456}t_{712}  }
\times
 {  \spa{m_1}.{\ell_1}\spa{m_1}.{\ell_2}
\over \spa2.{\ell_2}\spa{\ell_1}.7 }
\times
{ \la \ell_1 | P_{712} P_{456} | 5 \ra \la \ell_2 | P_{712} P_{456} | 5 \ra  \over
 \la \ell_1 | P_{712} | 3 ]  \la \ell_2 | P_{712} P_{456} | 4 \ra 
  }.
\cr}
\equn
$$
This can be recognised as a ${\cal H}_4$ canonical form and thus yields a contribution to the bubble coefficient of,
$$
\eqalign{
& 
-{\la m_1|P_{712}P_{456}|5\ra^2
\over 
 \spa7.1 \spa1.2\spa4.5\spa5.6 \BRi36{456}t_{456}t_{712}  }
\cr &
\hskip 2.0 truecm \times
H_4[ 2,7, P_{712}|3] , P_{712}P_{456}|4\ra ; m_1,m_1, P_{712}P_{456}|5\ra,P_{712}P_{456}|5\ra ; P_{712} ].
\cr}
\equn
$$
Similarly the $T_{2}$ and $T_{3}$ terms give a contribution,
$$
\eqalign{
 &
{1
\over  \spa7.1 \spa1.2 \spa3.4\spb5.6}
G_4[ 2,7, P_{34}|5] , P_{712}P_{56}|4\ra ; m_1,m_1,3,Y_{B2},Y_{B2} ; 6 ; P_{34} ; P_{712}  ]
\cr
+&
{ [4|P_{456}|m_1\ra^2
\over \spa7.1 \spa1.2 \spb3.4\spb4.5 \BRi36{345}
t_{345} }
H_4 [ 2,7,6, P_{345}|5] ; m_1,m_1, P_{345}|4] ,P_{345}|4]; P_{712}  ].
\cr}
\equn
$$

We define the $C_B$ function by generalising the result of this specific cut: 
$$
\eqalign{
C_B&[a,b,c,d,e,f,g; m_1]\equiv
\cr
&-{\la m_1|P_{gab}P_{def}|e\ra^2
\over 
 \spa{g}.{a} \spa{a}.{b}\spa{d}.{e}\spa{e}.{f} [c|P_{def}|f\ra t_{def}t_{gab}  }
\cr &
\hskip 2.0 truecm \times
H_4[ b,g, P_{gab}|c] , P_{gab}P_{def}|d\ra ; m_1,m_1, 
P_{gab}P_{def}|e\ra,P_{gab}P_{def}|e\ra ; P_{gab}] 
\cr
+&{1
\over  \spa{g}.a \spa{a}.b \spa{c}.d\spb{e}.f}
G_4[ b,g, P_{cd}|e] , P_{gab}P_{ef}|d\ra ;  m_1,m_1,c,Y_{B2},Y_{B2} ; f; P_{cd}; P_{gab} ]
\cr
+&
{ [d|P_{cde}|m_1\ra^2
\over \spa{g}.a \spa{a}.b \spb{c}.d\spb{d}.e [c|P_{cde}|f\ra
t_{cde} }
H_4 [ b,g,f, P_{cde}|e] ; m_1,m_1, P_{cde}|d] ,P_{cde}|d]; P_{gab} ],
\cr}
\equn
$$
with,
$$
| Y_{B2} \ra =
\spb{f}.{e}\spa{e}.c |m_1\ra +
\spb{f}.g\spa{m_1}.c |g\ra+
\spb{f}.a\spa{m_1}.c |a\ra+
\spb{f}.b\spa{m_1}.c |b\ra.
\equn
$$

Five of the bubble coefficients can be expressed in terms of the $C_B$ function:
$$
\eqalign{
c_7^C = C_B[1,2,3,4,5,6,7;1],
\;& \;
c_6^C = -C_B[7,6,5,4,3,2,1;1],
\;  \;
\cr
c_5^C = C_B[6,7,1,2,3,4,5;5],
\;& \;
c_4^C = -C_B[5,4,3,2,1,7,6;5],
\;\;
\cr
c_6^A  = C_B[7,1,2,3,4,5,6;1].\; &
\cr}
\equn
$$

We define three further functions in this class, $C_A$, $C_C$ and $C_D$;  
$$
\eqalign{C_A[a,&b,c,d,e,f,g; m_1]  \equiv
\cr &{ \la m_1| P_{gab}P_{cd}| e\ra^2\over\spa{c}.{d}\spa{d}.{e} [ f|P_{de}| c\ra \spa{g}.{a} \spa{a}.{b}t_{cde} t_{gab} }H_3[ b, g,   P_{gab}|f]  ; m_1, m_1, P_{gab} P_{cd}| e\ra ; P_{gab} ]\cr&-{ [d|P_{ef}|m_1\ra^2\over \spa{g}.{a}\spa{a}.{b}\spb{d}.{e}\spb{e}.{f} [f|P_{de}|c\ra t_{def} } H_3[ b, c, g ; m_1, m_1,P_{ef} |d]; P_{gab}], 
\cr}
\equn
$$
$$
\eqalign{C_C [a,&b,c, d,e,f,g; m_1]  
\equiv\cr& 
{ \spb{d}.{e}^3 \spa{m_1}.{f}^2\over            t_{cde} \spb{c}.{d} 
[c|P_{de}| f\ra \spa{g}.{a}\spa{a}.{b}}H_3[g, b, P_{cd}|e] ; f, m_1, m_1;  P_{gab}]\cr-&{ \la  m_1 | P_{gab} P_{de}| f\ra^2\over              t_{def} t_{gab} \spa{d}.{e} \spa{e}.{f} [c|P_{de}| f\ra              \spa{g}.{a}\spa{a}.{b}}\cr & \hskip 2.0truecm \times\overline{H}_4 [c,  P_{ef}|d\ra , P_{ga}|b\ra,P_{ab}|g\ra ; P_{de}|f\ra ,         P_{de}|f\ra ,  P_{gab}|m_1 \ra , P_{gab}|m_1 \ra           ;P_{gab}]\cr+&{    1\over \spb{e}.{f} \spa{c}.{d} \spa{g}.{a} \spa{a}.{b}  } \cr & \hskip 2.0truecm \times G_5[ P_{cd}|e] ,  P_{gab} P_{ef}|d\ra , P_{gab}|f], b, g ;     c, m_1, m_1, P_{gab}|e], Y_C, Y_C;    e; P_{cd} ;  P_{gab}], 
\cr}
\equn$$
$$
\eqalign{
C_D [a,&b,c,d,e,f,g; m_1]  \equiv
\cr &
{ \spa{d}.{e}^3  [f|P_{gab}|m_1\ra^2\over t_{cde} t_{gab}  \spa{c}.{d}  [f|P_{de}|c\ra \spa{g}.{a} \spa{a}.{b} }H_3[ b, g, P_{gab}P_{cd}|e\ra ; m_1, m_1,  P_{gab}|f]  ; P_{gab} ]\cr-&{ [f|P_{de}|m_1\ra^2\over t_{def}\spb{d}.{e}\spb{e}.{f} [f|P_{de}|c\ra \spa{g}.{a} \spa{a}.{b}  }H_4[ b, c, g, P_{ef}|d]; m_1, m_1, P_{de}|f], P_{de}|f]; P_{gab} ]\cr-&{ 1\over\spa{e}.{f}\spb{c}.{d}\spa{g}.{a}\spa{a}.{b} }G_5[ f, b, g, P_{ef}|d], P_{gab}P_{cd}|e\ra;     e, e, m_1, m_1, Y_D, Y_D ;    c;P_{ef}; P_{gab}], 
\cr}
\equn$$
where,
$$
\eqalign{
|Y_C\ra =& \spb{e}.{f}\spa{f}.{c}|m_1\ra
           +\spa{m_1}.{c}(\spb{e}.{g}|g\ra+\spb{e}.{a}|a\ra+\spb{e}.{b}|b\ra),
\cr
|Y_D\ra =&-\spb{c}.{d}\spa{d}.{e}|m_1\ra
           +\spa{e}.{m_1}(\spb{c}.{g}|g\ra+\spb{c}.{a}|a\ra+\spb{c}.{b}|b\ra).
\cr}
\equn
$$

\def\charskip{\hskip 14pt}
The remaining $c^X_i$ coefficients are then:
$$
\eqalign{
c_1^A=
 C_0[2,3,4,5,6,7,1;3,4],
\;\; &
c_2^A=
C_0[3,4,5,6,7,1,2;3,1],
\cr
c_3^A=
C_A[4,5,6,7,1,2,3;4],
\;\; &
c_4^A=
C_D[5,6,7,1,2,3,4;4],
\cr
c_7^A=&
 C_0[1,2,3,4,5,6,7;7,4],
\cr
&\cr
c_1^B=
C_0[2,3,4,5,6,7,1;3,5],
\;\; &
c_2^B=
C_C[3,4,5,6,7,1,2;2],
\cr
c_3^B=
 C_A[4,5,6,7,1,2,3;5],
\;\; &
c_4^B=
 C_D[5,6,7,1,2,3,4;5],
\cr
c_5^B=
 -C_A[6,5,4,3,2,1,7;5],
\;\; &
c_6^B=
-C_C[7,6,5,4,3,2,1;1],
\cr
c_7^B=&
-C_0[1,7,6,5,4,3,2;7,5],
\cr
&\cr
c_1^C=
 C_0[2,3,4,5,6,7,1;5,2],
\;\;&
c_2^C=
C_C[3,4,5,6,7,1,2;3],
\cr
c_3^C=&
C_0[4,5,6,7,1,2,3;4,1].
\cr}
\equn
$$

\subsection{$D_X$ Functions} 
\label{DAbody}

The $D_X$ functions arise as the coefficients of $\Ibub (s_{ab})$ when we consider cuts of the form,
$$
\int \dlips\; A^{\rm tree}(-\ell_1,a,b,\ell_2)  \times A^{\rm tree} (-\ell_2,c,d,e,f,g,\ell_1) \to D[a,b,c,d,e,f,g] \; .
\equn
$$
In order to obtain non-vanishing $\NeqOne$ and scalar contributions, legs $a$ and $b$ must be of opposite helicity.  
For the $\NeqOne$ contribution the two helicity configurations for $a$ and $b$ are trivially related:
$$
D[a^+,b^-,c,d,e,f,g] =-D[b^-,a^+,c,d,e,f,g] \;.
\equn
$$
Consequently, the number of independent $D_X$ functions corresponds to the number of independent
helicity configurations for the legs $c,d,e,f,g$ that contain two negative and
three positive helicities. There are six such configurations:  $(--+++)$, $(-+-++)$, $(-++-+)$, $(-+++-)$, 
$(+--++)$ and $(+-+-+)$.   

To evaluate these cuts we need the explicit forms of the seven-point
NMHV tree amplitudes where two external states are 
scalars or fermions (given explicitly in Appendix~\ref{SevenPointTree}).  Using these forms for the tree amplitude 
$A^{\rm tree}(\ell_2^h,c^-,d^-,e^+,f^+,g^+,\ell_1^{-h})$, we express the cut of the first of these helicity configurations in terms of canonical forms and obtain, 
$$
\eqalign{
D_A&[a,b,c,d,e,f,g]=\cr&{[e|P_{cde}|a\ra^2\over\spa{a}.{b}\spa{f}.{g}\spb{c}.{d}\spb{d}.{e}[c|P_{cde}|f\ra t_{cde}}H_2[b,g;a,P_{cde}|e];P_{ab}]\cr&-{\spa{f}.{d}\over\spb{a}.{b}\spa{d}.{e}\spa{e}.{f}\spa{f}.{g}[c|P_{de}|f\ra }\cr&
\overline{H}_4[a,c,P_{cde}|f\ra ,P_{ab}P_{abc}P_{de}|f\ra ;b,g,X_A,X_A;P_{ab}]\cr&-{1\over\spb{a}.{b}\spa{d}.{e}\spa{e}.{f}\spa{f}.{g}[c|P_{de}|f\ra }\cr&
\overline{H}_5[a,P_{ab}|g\ra ,c,P_{cde}|f\ra ,P_{ab}P_{abc}P_{de}|f\ra ;b,P_{ab}|f\ra ,P_{ef}|d\ra ,X_A,X_A;P_{ab}]\cr&-{\spa{c}.{d}^4\spb{g}.{b}^2\over\spb{a}.{b}\spa{c}.{d}\spa{d}.{e}\spa{e}.{f}[g|P_{ab}|c\ra t_{gab}}
H_2[b,P_{ab}P_{gab}|f\ra ;a,P_{ab}|g]; P_{ab}]\cr&-{\spa{a}.{c}^2[g|P_{abc}|d\ra^4\over\spa{a}.{b}\spa{d}.{e}\spa{e}.{f}[g|P_{ab}|c\ra [g|P_{abc}|d\ra t_{abc}t_{def}}
H_2[b,P_{abc}P_{de}|f\ra ;a,c; P_{ab}]
,\cr}
\equn
$$
where,
$$
\eqalign{
|X_A]= &\spa{f}.{d}\spb{b}.{a}\spa{a}.{g}|g]     +\spa{f}.{d}\spb{b}.{a}\spa{a}.{f}|f]      -\spa{a}.{f}\spb{b}.{a}\spa{d}.{e}|e]\cr &      +(\spa{f}.{g}\spb{g}.{c}\spa{c}.{d}      +[c|P_{abc}|f\ra\spa{c}.{d}      +\spa{f}.{d}s_{ab})|b].
\cr}
\equn
$$


\def\charskip{\hskip 14pt}
For the other helicity configurations we define $D_{B,C,D,E,F}$ in a similar fashion. The explicit forms
of these are given in appendix~\ref{DXap}. In terms of these we have,
$$
\eqalign{
d_2^A=D_D[2,3,4,5,6,7,1],
\;\;\;
&d_3^A=
-D_A[4,3,2,1,7,6,5],
\cr
d_4^A=
-D_E[5,4,3,2,1,7,6],
\;\;\;
&d_7^A= D_B[7,1,2,3,4,5,6],
\cr
& \cr
d_2^B=-D_C[2,1,7,6,5,4,3],
\;\;\;
&d_4^B=D_E[4,5,3,2,1,7,6],
\cr
d_5^B= -D_E[6,5,7,1,2,3,4],
\;\;\;
&d_7^B= D_C[1,2,3,4,5,6,7],
\cr
& \cr
d_1^C=-D_B[2,1,3,4,5,6,7],
\;\;\;
&d_2^C=  D_C[3,1,7,6,5,4,2],
\cr
d_3^C= -D_C[3,5,6,7,1,2,4],
\;\;\;
&d_4^C=D_B[4,5,3,2,1,7,6],
\cr
d_5^C= D_F[5,6,7,1,2,3,4],
\;\;\;
&d_7^C= -D_F[1,7,6,5,4,3,2].
\cr}
\equn
$$

For each amplitude we have checked at explicit kinematic points that these bubble coefficients satisfy,
$$
\sum_i c_i +\sum_j d_j = { A^{\rm tree} }\, .
\equn
$$
These conditions ensure that each amplitude has the correct $1/\eps$ IR singularity~\cite{Kunszt:1994mc}.

\section{Canonical Basis for Triangle Coefficients from Triple Cuts}

Generalisations of unitarity can be used to determine the coefficients of triangle and box functions simply. 
If we consider a triple cut, 
$$
\eqalign{
C_3\,=\,
-\int d^4 \ell \delta(\ell_0^2)  \delta(\ell_1^2) \delta(\ell_2^2) 
& \sum  A\Big((-\ell_0), m, \cdots, j-1, (\ell_1) \Big)
\cr
&\hspace{-3cm}
\,\times\,
A\Big((-\ell_1), {j} ,\cdots, {l-1}, (\ell_2) \Big)
\,\times\,
A\Big((-\ell_2), {l} ,\cdots, {m-1},  (\ell_0) \Big)\,,
\cr}
\hspace{-1cm}
\equn
$$ 
The minus sign in this equation is for when using colour-ordered partial amplitudes as normalised in eqn.~(\ref{ColourDecomposition}). 
This has contributions from the discontinuities of a single triangle and several box functions,
$$
C_3=\left(
\sum_{i\in \cal C''}\, a_i\, I_4^i +b_j\, I^j_3 
\right) {\Big|}_{\rm Disc} \ . 
\equn\label{DecompDisc}
$$
The information in this cut may be used to determine the triangle coefficient $b_j$. 

One of the advantages of the supersymmetric decomposition of gluonic amplitudes is that the one and two-mass triangle coefficients
need not be explicitly computed: for the $\NeqFour$ contributions they are absent 
while for the $\NeqOne$ and scalar contributions they are tied to the box coefficients and can
be incorporated into the truncated box functions.  
Consequently this section will focus on the case where all three masses of the triangle are non-zero.
Consider a physical triple cut in an amplitude where all three corners
are non-null
 \vspace{0.6cm}
\begin{center}
\begin{picture}(40,60)(-20,50)
\Line(30,30)(70,40)
\Line(30,30)(70,20)
\SetWidth{2}
\Line(30,30)(60,50)
\Line(30,30)(60,10)
\SetWidth{1}
\Line(30,30)(-30,30)
\Line(-30,30)(0,75)
\Line(30,30)(0,75)
\Line(-30,30)(-70,40)
\Line(-30,30)(-70,20)
\SetWidth{2}
\Line(-30,30)(-60,50)
\Line(-30,30)(-60,10)
\SetWidth{1}
\Text(-60,30)[]{$\bullet$}
\Line(0,75)(-10,105)
\Line(0,75)(10,105)
\SetWidth{2}
\Line(0,75)(-20,95)
\Line(0,75)(20,95)
\Text(0,100)[]{$\bullet$}
\Text(57,41)[]{$\bullet$}
\Text(57,18)[]{$\bullet$}
 \SetWidth{1}
\DashCArc(45,20)(40,100,190){4}
\DashCArc(-50,20)(40,-10,80){4}
\DashCArc(00,100)(40,220,320){4}
\CCirc(30,30){8}{Black}{Purple}
\CCirc(-30,30){8}{Black}{Purple}
\CCirc(0,75){8}{Black}{Purple}
\Text(0,10)[]{$\ell_0$}
\Text(-30,65)[]{$\ell_1$}
\Text(30,65)[]{$\ell_2$}
\Text(80,30)[]{$K_3$}
\Text(-80,30)[]{$K_1$}
\Text(0,115)[]{$K_2$}
\end{picture}
\end{center}
\vspace{1.7cm}

\begin{figure}[h]
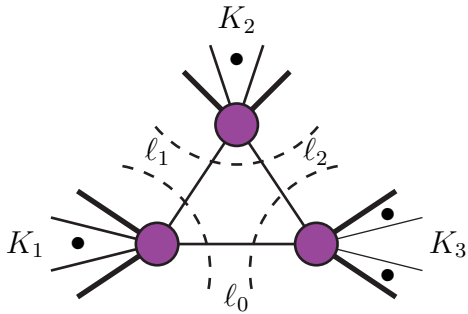


\caption{The triple cut of an amplitude}
\label{FIGTripleCut}\end{figure}

As
the momentum invariants, $K_1\equiv k_{m}+k_{m+1}+\cdots +k_{j-1}$ etc, are all
non-null, there exist kinematic regimes is which the integration has
non-vanishing support for real loop momentum.

In this section we present the contributions of 
various canonical forms to the coefficient of the three-mass triangle integral function. 
As before we build our canonical forms from a simple starting point.  
Consider,
$$
{\cal E}_1( a ; b) 
\equiv
{ 
\spa{\ell_0}.{b} 
\over
\spa{\ell_0}.{a}  
}
=  {  [a |\ell_0 | b \ra \over (\ell_0+k_a)^2  }.
\equn
$$
We chose to
manipulate this as if it were a covariant integral. This means effectively
replacing, 
$$
\int \dlips \longrightarrow  
 \int { d^D\ell  \over \ell_0^2 \ell_1^2 \ell_2^2 },
\equn
$$
then evaluating the covariant integral 
only keeping the coefficient of  the three-mass triangle in the  result.  

The covariant integral is a linear box function.  
Evaluating this  we find 
the following contribution to the three-mass triangle coefficient:
$$
E_1[a;b]=
-{  \la a |[ K_1,K_2 ] | b \ra
\over 2 \la a | K_1 K_2 | a \ra  
}.
\equn
$$
As in eqn.~(\ref{PQdefnEQ}), we could write the denominator as,
$$
\la a | \hat K_1 \hat K_2 | a \ra  =\spa{a}.{\hat K_1}\spb{\hat K_1}.{\hat K_2} \spa{\hat K_2}.{a} \, ,
\equn
$$
where $\hat K_i$ are the null linear combinations of $K_1$ and $K_2$. These linear combinations
involve irrational coefficients as in eqn.(\ref{PQdefnBEQ}). 
The case where $\spa{a}.{\hat K_i} =0$ must be treated as a special case. This does not arise when $a$ denotes an 
external momentum but may
arise in the derivations of more complicated canonical forms. 
When ${a}={\hat K_3}$, we have,
$$
{\cal E}_1^x =  {\spa{b}.{\ell}
\over \spa{\hat K_3}.\ell }
\longrightarrow  E_1^x[ K_3,b]=-
{ \la b | K_1 | K_3 ] \over 2 \la K_3 | K_1 | K_3 ] }.
\equn
$$

More complicated forms can readily be generated from this simple starting point. A summary of these 
canonical forms is given in appendix~\ref{CanonTripApp}.  
As an example of the use of these canonical forms, the general expression for a NMHV three-mass 
triangle coefficient is given in the next section. 

\section{Example: $n$-point Three-Mass Triangles for $\NeqOne$ and Scalar Contributions to NMHV Amplitudes }
\label{gen3masstrisection}

As an example of using the canonical forms for the three mass triangle let us evaluate the general form 
for the three-mass triangle for NMHV amplitudes. The $\NeqOne$ contribution was 
previously presented in~\cite{BjerrumBohr:2007vu}.

For both the chiral $\NeqOne$ and scalar contributions, the only three-mass triangles which appear in the NMHV amplitude 
have exactly one negative helicity on each corner. Consider such an integral function with the following labelling:  
 
\begin{center}
\begin{picture}(40,60)(-20,50)
\Line(30,30)(70,40)
\Line(30,30)(70,20)
\SetWidth{2}
\Line(30,30)(60,50)
\Line(30,30)(60,10)
\SetWidth{1}
\Line(30,30)(-30,30)
\Line(-30,30)(0,75)
\Line(30,30)(0,75)
\Line(-30,30)(-70,40)
\Line(-30,30)(-70,20)
\SetWidth{2}
\Line(-30,30)(-60,50)
\Line(-30,30)(-60,10)
\SetWidth{1}
\Text(-60,30)[]{$\bullet$}
\Line(0,75)(-10,105)
\Line(0,75)(10,105)
\SetWidth{2}
\Line(0,75)(-20,95)
\Line(0,75)(20,95)
\Text(0,100)[]{$\bullet$}
\Text(57,41)[]{$\bullet$}
\Text(57,18)[]{$\bullet$}
 \SetWidth{1}
\Text(80,35)[]{$K_3$}
\Text(-80,30)[]{$K_1$}
\Text(-5,115)[]{$K_2$}

\Text(25,100)[]{$u_2$}
\Text(-27,102)[]{$f_2$}
\Text(15,115)[]{$m_2^-$}

\Text(70,10)[]{$u_3$}
\Text(70,55)[]{$f_3$}
\Text(85,25)[]{$m_3^-$}

\Text(-70,55)[]{$u_1$}
\Text(-70,8)[]{$f_1$}
\Text(-85,15)[]{$m_1^-$}

\end{picture}
\end{center}
\vspace{0.7cm}
\begin{figure}[h]
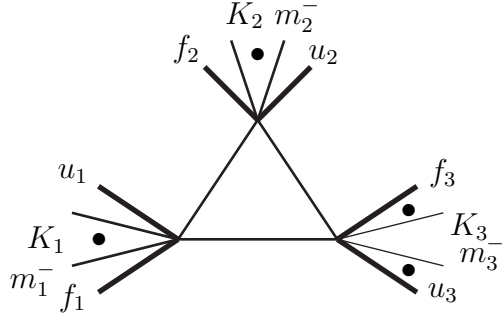


\caption{The three mass triangle integral function appearing in the NMHV 
amplitudes indicating the labeling of the legs.}
\label{FIGTripleIntegral}
\end{figure}

\vskip 0.5 truecm

First we take the $\NeqOne$ case.
Evaluating this using the triple cut of fig.~\ref{FIGTripleCut} the triple cut integrand is,
$$
\sum_h
A(-\ell_0^h, \cdots , m_1^- , \cdots , \ell_1^{-h} )
\times
A(-\ell_1^h, \cdots , m_2^- , \cdots , \ell_2^{-h} )
\times
A(-\ell_2^h, \cdots , m_3^- , \cdots , \ell_0^{-h} ),
\equn
$$
where $\ell_{1}=\ell_{0} -K_1$ etc. 
The summation is over the $\NeqOne$ chiral multiplet.  The effect of this
summation is to give the scalar contribution times a $\rho$-factor,
$$
A(-\ell_0^0, \cdots , m_1^- , \cdots , \ell_1^{0} )
\times
A(-\ell_1^0, \cdots , m_2^- , \cdots , \ell_2^{0} )
\times
A(-\ell_2^0, \cdots , m_3^- , \cdots , \ell_0^{0} )
\times \rho,
\equn$$
where, 
$$
\eqalign{
\rho
&=
-{ (\spa{m_1}.{\ell_1}\spa{m_2}.{\ell_2}\spa{m_3}.{\ell_0}-\spa{m_1}.{\ell_0}\spa{m_2}.{\ell_1}\spa{m_3}.{\ell_2} )^2
\over 
 \spa{m_1}.{\ell_1}\spa{m_2}.{\ell_2}\spa{m_3}.{\ell_0}\spa{m_1}.{\ell_0}\spa{m_2}.{\ell_1}\spa{m_3}.{\ell_2} }
\cr
&=
-{\spa{\ell_0}.X^2  
\over 
 \spb{\ell_1}.{\ell_2}^2 \spa{m_1}.{\ell_1}\spa{m_2}.{\ell_2}\spa{m_3}.{\ell_0}\spa{m_1}.{\ell_0}\spa{m_2}.{\ell_1}
\spa{m_3}.{\ell_2} }\, ,
\cr}
\equn
$$
and,
$$
|X\ra = 
|m_1\ra \la m_3 | K_3K_2|m_2\ra 
+|m_3\ra  \la m_1 | K_1K_2|m_2\ra\,.
\equn
$$
The cut is then,
$$
\eqalign{
&-{ \spa{m_1}.{\ell_0}\spa{m_1}.{\ell_1}
\over \spa{\ell_0}.{f_1}\la f_1 \cdots u_1 \ra \spa{u_1}.{\ell_1}\spa{\ell_1}.{\ell_0} }
\times
{ \spa{m_2}.{\ell_1}\spa{m_2}.{\ell_2}
\over \spa{\ell_1}.{f_2}\la  f_2 \cdots u_2 \ra \spa{u_2}.{\ell_2}\spa{\ell_2}.{\ell_1} }
\cr & \hskip 7.0truecm 
\times
{ \spa{m_3}.{\ell_2}\spa{m_3}.{\ell_0}
\over \spa{\ell_2}.{f_3}\la f_3 \cdots u_3 \ra \spa{u_3}.{\ell_0}\spa{\ell_0}.{\ell_2} }
\times
{ \spa{\ell_0}.X^2 \over \spb{\ell_1}.{\ell_2}^2 } 
\cr
&= C_0
\times 
{ \spa{m_1}.{\ell_0}\spa{m_1}.{\ell_1}
\over \spa{\ell_0}.{f_1} \spa{u_1}.{\ell_1} }
\times
{ \spa{m_2}.{\ell_1}\spa{m_2}.{\ell_2}
\over \spa{\ell_1}.{f_2} \spa{f_2}.{\ell_2} }
\times
{ \spa{m_3}.{\ell_2}\spa{m_3}.{\ell_0}
\over \spa{\ell_2}.{f_3}\spa{u_3}.{\ell_0} }
\times { \spa{X}.{\ell_0}^2  \over \la \ell_0 | \ell_1\ell_2 | \ell_0 \ra },
\cr}
\equn
$$
where,
$$
C_0 = - {  \spa{u_1}.{f_2} \spa{u_2}.{f_3}  \spa{u_3}.{f_1}
\over \prod_i \spa{i}.{i+1} K_2^2  }\,.
\equn
$$

This can be turned into a function of $\ell_0$ only:
$$
C_0
\times 
{ \prod_{|y\ra \in T_1} \spa{\ell_0}.{y}  
\over \prod_{|x\ra \in S} \spa{\ell_0}.{x}  }
\times { 1  \over \la \ell_0 | K_1K_3 | \ell_0 \ra }\,,
\equn
$$
where,
$$
\eqalign{
S &=\{  |f_1\ra, |u_3\ra,  K_3K_2|f_2\ra , K_3K_2|u_1\ra , K_1K_2|f_3\ra , K_1K_2|u_2\ra \},
\cr
T_1 &=\{ |m_1\ra ,  |m_3\ra ,   K_3K_2|m_1\ra,  K_3K_2|m_2\ra , K_1K_2|m_2\ra, K_1K_2|m_3\ra , |X\ra , |X\ra \},
\cr}
\equn
$$
and we have used,
$$
{ \spa{\ell_1}.a  \over \spa{\ell_1}.b }
= { \la \ell_0 | \ell_2 \ell_1 | a \ra 
\over
 \la \ell_0 | \ell_2 \ell_1 | b \ra  }
=
{ \la \ell_0 | K_3 K_2 | a \ra 
\over
 \la \ell_0 | K_3 K_2 | b \ra  },
\;\;\;\;\;\;
{ \spa{\ell_2}.a  \over \spa{\ell_2}.b }
= { \la \ell_0 | K_1 K_2 | a \ra 
\over
 \la \ell_0 | K_1 K_2 | b \ra  }.
\equn
$$
This is precisely the canonical form ${\cal J}_n^0$ with $n=6$ as defined in appendix~\ref{CanonTripApp}. So
the three mass triangle coefficient is precisely,
$$
b^{3m}_3  = C_0 \times J^0_6(  S ; T_1 ; K_i )\, .
\equn
$$
This general expression simplifies in many cases: if the $m_i$ coincide with any of the $u_i$ or 
$f_i$, the $J^0_6$ function simplifies to $J^0_n$ with 
$n < 6$.

The scalar case is more complicated. The 
cut integrand is,
$$
\eqalign{
&{ \spa{m_1}.{\ell_0}^2\spa{m_1}.{\ell_1}^2
\over \spa{\ell_0}.{f_1}\la f_1 \cdots u_1 \ra \spa{u_1}.{\ell_1}\spa{\ell_1}.{\ell_0} }
\times
{ \spa{m_2}.{\ell_1}^2\spa{m_2}.{\ell_2}^2
\over \spa{\ell_1}.{f_2}\la  f_2 \cdots u_2 \ra \spa{u_2}.{\ell_2}\spa{\ell_2}.{\ell_1} }
\cr & \hskip 7.0truecm 
\times
{ \spa{m_3}.{\ell_2}^2\spa{m_3}.{\ell_0}^2
\over \spa{\ell_2}.{f_3}\la f_3 \cdots u_3 \ra \spa{u_3}.{\ell_0}\spa{\ell_0}.{\ell_2} }
\cr
&= C_0 \times
{ \spa{m_1}.{\ell_0}^2\spa{m_1}.{\ell_1}^2
\over \spa{\ell_0}.{f_1} \spa{u_1}.{\ell_1} }
\times
{ \spa{m_2}.{\ell_1}^2\spa{m_2}.{\ell_2}^2
\over \spa{\ell_1}.{f_2} \spa{f_2}.{\ell_2} }
\times
{ \spa{m_3}.{\ell_2}^2\spa{m_3}.{\ell_0}^2
\over \spa{\ell_2}.{f_3}\spa{u_3}.{\ell_0} }
\times {\spb{\ell_1}.{\ell_2}^2  \over \la \ell_0 | \ell_1\ell_2 | \ell_0 \ra }
\cr
&= C_0 \times
{ \spa{m_1}.{\ell_0}^2\spa{m_3}.{\ell_0}^2 
\over \spa{\ell_0}.{f_1} \spa{u_1}.{\ell_0}   }
\times
{  \spa{m_1}.{\ell_1}^2
\over \spa{\ell_1}.{f_2} \spa{u_2}.{\ell_1}  }
\times
{ \spa{m_3}.{\ell_2}^2
\over \spa{f_3}.{\ell_2} \spa{\ell_2}.{f_3}}
\times {\spa{m_2}.{\ell_2}^2 \spb{\ell_1}.{\ell_2}^2  \spa{m_2}.{\ell_1}^2 \over \la \ell_0 | K_2K_3 | \ell_0 \ra } 
\cr
&= C_0 \times
{ \prod_{|y\ra \in T_2} \spa{\ell_0}.{y}  
\over \prod_{|x\ra \in S} \spa{\ell_0}.{x}  }
\times {\spa{m_2}.{\ell_2}^2 \spb{\ell_1}.{\ell_2}^2  \spa{m_2}.{\ell_1}^2 \over \la \ell_0 | K_2K_3 | \ell_0 \ra }, 
\cr}
\equn
$$
where,
$$
\eqalign{
T_2 &=\{ |m_1\ra , |m_1\ra, |m_3\ra, |m_3\ra , K_1K_2|m_3\ra ,K_1K_2|m_3\ra , K_3K_2|m_1\ra, K_3K_2|m_1\ra \}.
\cr}
\equn
$$
Splitting the pole and defining $T' =T_2-\{ |m_1\ra,|m_1\ra,|m_3\ra \}$, the cut integrand becomes,
$$
\eqalign{
&C_0
{ \prod_{|y\ra \in T_2} \spa{\ell_0}.{y}  
\over \prod_{|x\ra \in S} \spa{\ell_0}.{x}  }
\times {  \la m_2 | \ell_1 \ell_2 | m_2 \ra^2  \over \la \ell_0 | K_2K_3 | \ell_0 \ra } 
\cr
=&
C_0
\sum_{|x\ra\in S} C_x { \spa{\ell_0}.{m_1} \spa{\ell_0}.{m_1} \spa{\ell_0}.{m_3}\over \spa{\ell_0}.{x} }
\times { ( \la m_2 | K_1 K_3 | m_2\ra  +\la m_2 | \ell_0 K_2 | m_2  \ra)^2  \over \la \ell_0 | K_2K_3 | \ell_0 \ra } 
\cr
\to &
C_0
\sum_{|x\ra\in S} C_x 
\biggl( \la m_2 | K_1 K_3 | m_2\ra^2  J^0_1( x; m_1 ,m_1,m_3   )
\cr
\null & \hskip  2.0truecm +2\la m_2 | K_1 K_3 | m_2\ra J^1_1(x; m_2 , m_1 ,m_1,m_3 ; K_2 | m_2\ra    )
\cr
\null & \hskip  2.0truecm + J^2_1(x ; m_2 , m_2 , m_1 ,m_1,m_3  ; K_2 | m_2\ra, K_2 | m_2\ra  )
\biggr), 
 \cr}
\equn
$$
where, 
$$
C_x={ \prod_{|y\ra \in T'} \spa{y}.x \over 
 \prod_{|z\ra \in S-\{ |x\ra \}} \spa{z}.x 
}. 
\equn
$$

This is a general, explicit,  rational form of the $n$-point NMHV three-mass triangle coefficient. 

\section{Summary}

We have presented an implementation of the Unitarity method and applied it to the computation of the seven-point one-loop $\NeqOne$ amplitudes. 
Once the canonical forms are derived, the method is algebraic. 
In many ways our canonical form approach is equivalent to alternate methods however it naturally applies to trees written in the spinor helicity formalism 
and produces results which are manifestly rational.  

\vskip 0.5 truecm 

\noindent
{\bf Acknowledgements: }
This research was supported by the Science and Technology
Facilities Council of the UK.

\appendix

\section{Integral Functions} 
\label{IntegralFunctionsAp}
\def\tn#1#2{t^{[#1]}_{#2}}
\def\e{\eps}

The scalar box integral is,
$$
I_4 = -i \L4\pi\R^{2-\e} \,\int {d^{4-2\e}p\over \L2\pi\R^{4-2\e}}
\;{1\over p^2 \L p-K_1\R^2 \L p-K_1-K_2\R^2 \L p+K_4\R^2}\;,
\equn
$$
where $K_i$ is  the sum of the momenta of the legs attached to the $i$-th corner. 
If a single leg is attached then $K_i$ is null. The form of the integral depends upon the 
number of the $K_i$ which are non-null, $K_i^2\neq 0$. We often misname these {\it massive 
legs}. The integrals are functions of the non-zero $K_i^2$ and the invariants,
$$
S \equiv (K_1+K_2)^2 , \;\;\; T \equiv (K_2+K_3)^2 .
\equn
$$
For convenience, we always define these integrals taking leg $1$ as massless and leg $4$ as massive. 

It is convenient to define the scalar box function, $F$, 
$$
F(K_1,K_2,K_3,K_4) = -{2\sqrt{\mathop{\rm det} S}\over\rg} \,I_4,
\equn
$$
where,
$$
\rg =  {\Gamma(1+\e)\Gamma^2(1-\e)\over\Gamma(1-2\e)},
\equn
$$
and the symmetric $4\times4$
matrix $S$ has components ($i$, $j$ are ${\rm mod} 4$),
$$
S_{ij} = -{1\over2}\L K_i+\cdots+ K_{j-1}\R^2, \quad i\neq j;
\hskip 1truecm S_{ii} = 0.
\equn
$$
In terms of these variables, the relationships between the scalar box
functions and scalar box integrals are given by,
$$
\eqalign{
  I_{4}^{1{\rm m}} &=\ -2 \rg {\Fone{i} \over S T }
        \, , \;\;\;
 I_{4}^{2{\rm m}e}
=\ -2 \rg {\Feasy{r;i}
      \over ST  -K_2^2K_4^2  }\,, 
\cr
 I_{4}^{2{\rm m}h}
&=\ -2 \rg {\Fhard{r;i} \over ST  } , \;\;\;
I_{4}^{3{\rm m}}
=\ -2 \rg {\Fthree{r,r';i}
     \over ST -K_2^2 K_4^2   }\,.
 \cr}
\equn
 $$
\begin{figure}[h]
\centering
\resizebox*{6cm}{!}{\includegraphics{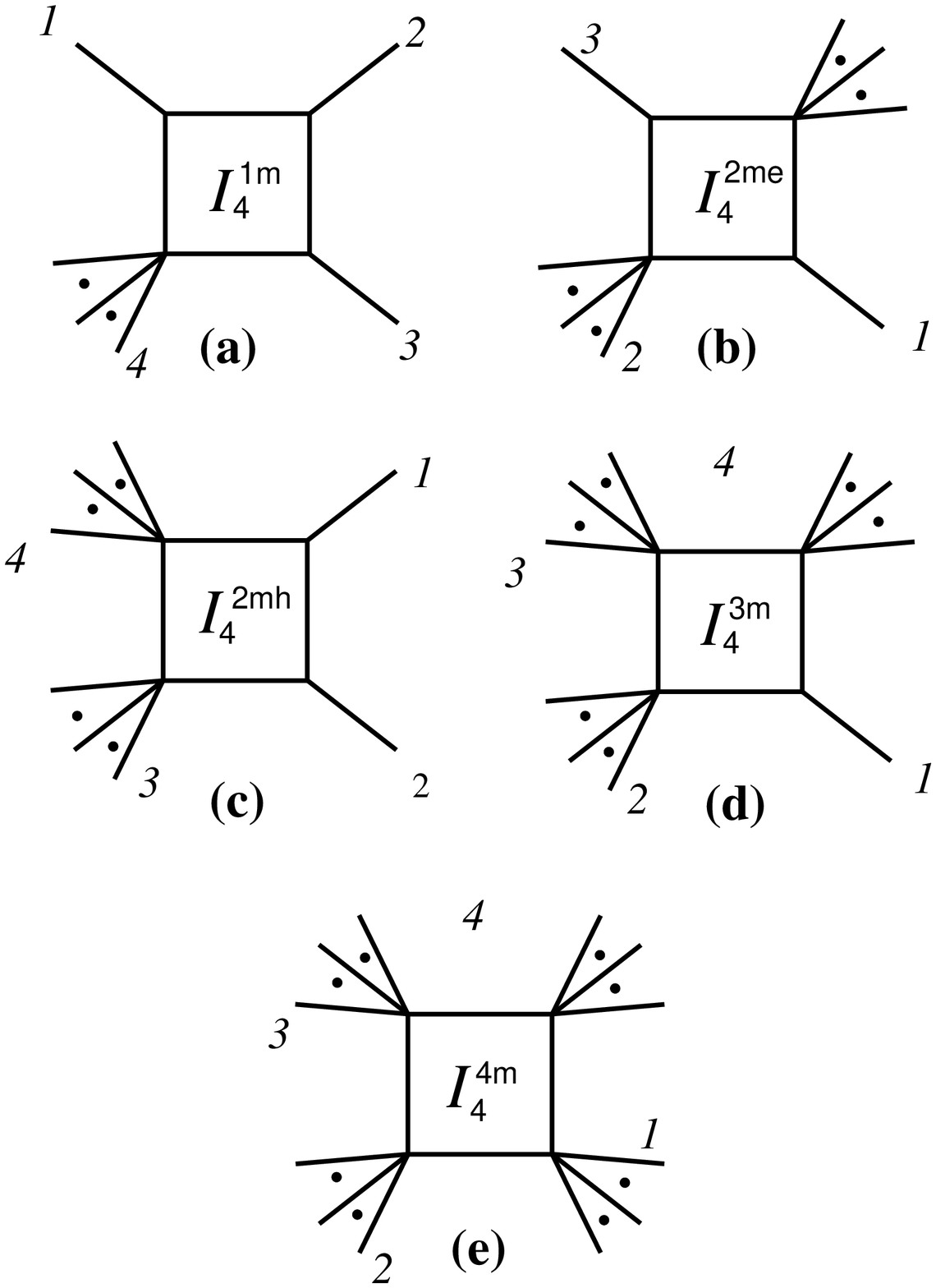}}
\caption{The different types of scalar box.}
\label{FIGboxtypes}
\end{figure}

With the labelling of legs shown in fig.~\ref{FIGboxtypes},
the scalar box functions expanded to ${\cal O}(\e^0)$
in the different cases reduce to,
$$
\eqalign{
  \Fone{i}
&=\ -{1\over\e^2} \Bigl[ (-S)^{-\e} +
(-T)^{-\e} - (-K_4^2)^{-\e} \Bigr] \cr
 &\ + \Li_2\left(1-{ K_4^2 \over S }\right)
  \ + \ \Li_2\left(1-{K_4^2 \over T}\right)
  \ +{1\over 2} \ln^2\left({ S  \over T}\right)\
+\ {\pi^2\over6}\ ,
\cr}
\equn
$$
$$
\eqalign{
  \Feasy{r;i}
&=\  - {1\over\e^2} \Bigl[ (-S)^{-\e} + (-T)^{-\e}
              - (-K_2^2 )^{-\e} - (-K_4^2 )^{-\e} \Bigr] \cr
&\ +\ \Li_2\left(1-{ K_2^2 \over S }\right)
 \ +\ \Li_2\left(1-{ K_2^2 \over T}\right)
 \ +\ \Li_2\left(1-{ K_4^2 \over S }\right)
\cr
&\
 \ +\ \Li_2\left(1-{ K_4^2 \over T }\right)
-\ \Li_2\left(1-{ K_2^2 K_4^2 
\over S T}\right)
   \ +\ {1\over 2} \ln^2\left({S/T}\right)\ , 
\cr 
}
\equn
$$
$$
\eqalign{ 
  \Fhard{r;i}
&=\ -{1\over\e^2} \Bigl[ (-S)^{-\e} + (-T)^{-\e}
              - (-K_3^2 )^{-\e} - (-K_4^2)^{-\e} \Bigr]\hskip 2.0truecm{}
\cr &
  \ -\ {1\over2\e^2}
    { (-K_3^2)^{-\e}(-K_4^2)^{-\e}
     \over (-S)^{-\e} }
  \ +\ {1\over 2} \ln^2\left({ S/T }\right)
\cr &
  \ +\ \Li_2\left(1-{ K_3^2\over T}\right)
  \ +\ \Li_2\left(1-{ K_4^2\over T }\right)
  \ , 
\cr
}
\equn
$$
$$
\eqalign{ 
  \Fthree{r,r';i}
&=\ -{1\over\e^2} \Bigl[ (-S )^{-\e} + (-T)^{-\e}
     - (-K_2^2)^{-\e}
     - (-K_3^2 )^{-\e}
     - (-K_4^2 )^{-\e} \Bigr] \cr
  &  \ -\ {1\over2\e^2}
   { (-K_2^2)^{-\e}(-K_3^2 )^{-\e} \over(-T)^{-\e} }
  \ -\ {1\over2\e^2}
    {(-K_3^2)^{-\e}(-K_4^2)^{-\e}
           \over (-S)^{-\e} }
      \ +\ {1\over2}\ln^2\left({S \over T}\right)
\cr
  &\ +\ \Li_2\left(1-{K_2^2\over S }\right)
   \ +\ \Li_2\left(1-{K_4^2\over T}\right)
  \ -\  \Li_2
\left(1-{K_2^2K_4^2\over ST }\right)
\ .  \cr }
\equn
$$
Our seven-point expressions do not need the explicit form of the four mass scalar box. 

At this point we must
discuss a suitable basis for expressing the amplitudes. We could use the basis
(\ref{DecompBasis}), however this is not the most efficient option. 
By choosing a suitable basis of box functions we can considerably simplify the structure of the triangle coefficients.

Triangle integral functions may have one, two or three massless legs: 
$I_{3}^{1\rm m}$, $I_{3}^{2\rm m}$, $I_{3}^{3\rm m}$. 
The one-mass triangle function depends only on the momentum invariant of the
massive leg $K_1$ and is,
$$
I_{3}^{1\rm m} = {\rg\over\e^2} (-K_1^2)^{-1-\eps} \ ,
\equn
$$ 
while 
the two-mass triangle function with non-null momenta $K_1$ and $K_2$ is,
$$
I_{3}^{2 \rm m} = {\rg\over\e^2}
{(-K_1^2)^{-\eps}-(-K_2^2)^{-\eps} \over  (-K_1^2)-(-K_2^2) }\ .
\equn
$$ 
Both of these integral functions contain $\ln(K^2)/\eps$ IR
singularities.  The key point is that the IR singularities of the
amplitudes must be~\cite{Kunszt:1994mc},
$$
A^{N=1\ \rm chiral}_{IR} =  {\cg \over \eps} \Atree \, ,
\;\;\;\;\;
A^{[0]}_{IR} =  {\cg \over 3 \eps} \Atree \, ,
\equn
$$ 
so the $\ln(K^2)/\eps$ singularities must cancel.  This
constraint effectively determines the coefficients of $I_{3}^{1\rm m}$
and $I_{3}^{2\rm m}$ in terms of the box coefficients.

Specifically the one- and two-mass triangles are linear combinations of
the set of functions,
$$
G(-K^2)= \rg { (-K^2)^{-\eps}  \over\e^2} \; ,
\equn
$$
with,
$$
\eqalign{
I^{1m}_3 &= G(-K_1^2)\;\; ,
\;\;
\cr
I^{2m}_3 &={ 1 \over (-K_1^2)-(-K_2^2) }
\left( G(-K_1^2)-G(-K_2^2  ) \right).
\cr}
\equn
$$
The $G(-K^2)$ are labeled by the independent momentum invariants
$K^2$ and in fact form an independent basis of functions, unlike the
one and two-mass triangles which are not all independent.

In practice we need never calculate the coefficients of the $G$ functions once we know the 
box coefficients. 
The only functions containing $\ln(s)/\eps$ terms are the box functions and $I_3^{1m}$ and $I_3^{2m}$ so,
$$
\sum a_i I_4^i|_{\ln(K^2)/\eps}  + b_G  { \ln(K^2) \over \eps }  
 = 
0 \, .
\equn
$$
This equation fixes the single $b_G$ in terms of the $a_i$.  
The simplest approach to implement this simplification is to express
the amplitudes in terms of truncated, finite ${\cal F}$-functions~\cite{BDDKa,Bidder:2004vx,BBCFsusyone}:
$$
\eqalign{
{\cal F}^{1m} &= F^{1m}
+{1 \over \eps^2} \left( { \mu^2 \over -S}\right)^{\eps} 
+{1 \over \eps^2}\left( { \mu^2 \over -T}\right)^{\eps}
-{1 \over \eps^2}\left( { \mu^2 \over -K_4^2 }\right)^{\eps} \, , 
\cr
{\cal F}^{2me}&
= F^{2me}
+{1 \over \eps^2} \left( { \mu^2 \over -S}\right)^{\eps} 
+{1 \over \eps^2}\left( { \mu^2 \over -T}\right)^{\eps}
-{1 \over \eps^2}\left( { \mu^2 \over -K_2^2 }\right)^{\eps}  
-{1 \over \eps^2}\left( { \mu^2 \over -K_4^2 }\right)^{\eps}   \, , 
\cr
{\cal F}^{2mh} &
= F^{2mh} 
+{ 1\over 2\eps^2} \left( { \mu^2 \over -S}\right)^{\eps} 
+{1 \over \eps^2}\left( { \mu^2 \over -T}\right)^{\eps}
-{1 \over 2\eps^2}\left( { \mu^2 \over -K_3^2 }\right)^{\eps}  
-{1\over 2\eps^2}\left( { \mu^2 \over -K_4^2 }\right)^{\eps}   \, , 
\cr
{\cal F}^{3m} & 
= F^{3m} 
+{ 1\over 2\eps^2} \left( { \mu^2 \over -S}\right)^{\eps} 
+{1 \over \eps^2}\left( { \mu^2 \over -T}\right)^{\eps}
-{1 \over 2\eps^2}\left( { \mu^2 \over -K_2^2 }\right)^{\eps}  
-{1\over 2\eps^2}\left( { \mu^2 \over -K_4^2 }\right)^{\eps} \, .  
\cr}
\equn
$$  
The $\NeqOne$ and scalar amplitudes can then be expressed as,
$$
\Aloop =  \sum  a_i {\cal F}^i + \sum b_j^{3m} I_3^{3m,j} + \sum c_k I_2^k  +R \, ,
\equn
$$
with no $I_3^{1m}$ or $I_3^{2m}$  present.

In transferring to this set of basis functions, 
the coefficients of the ${\cal F}^i$, $a_{\cal F}$, are simply related to the coefficients of the scalar box integrals, $a_I$, by, 
$$ 
\eqalign{
a^{1m}_{\cal F}  &= -{ 2 \over ST } a^{1m}_{I} \;,\;\;
a^{2me}_{\cal F}  = -{2 \over ST -K_2^2K_4^2 } a^{2me}_{I} \; ,
\cr
a^{2mh}_{\cal F}  &= -{2 \over ST } a^{2mh}_{I} \;,\;\;
a^{3m}_{\cal F}  = -{2 \over ST -K_2^2K_4^2 } a^{3m}_{I} \; .
\cr}
\equn
$$

The remaining integral functions are the three-mass triangles $I_3^{3m}$ as given, for example, in ref.~\cite{ThreeMassTriangle} and  the two-point bubble function,
$$
I_2(P)= {1 \over \eps} +2 -\ln(-P^2)\; .
\equn
$$
\section{Canonical Forms for Triple Cuts}
\label{CanonTripApp}

In this appendix the contributions of various canonical forms to the three-mass triangle coefficient are given. 
The triple cut is labelled as in figure~\ref{FIGTripleCut}.
$$
\eqalign{
{\cal E}_1 = & {\spa{b}.{\ell}
\over \spa{a}.\ell }
\longrightarrow  E_1[a;b; \{K_j\}]=-
{\la a | [ K_1,K_3]|b\ra \over 2\la a | K_1K_3|a\ra  }
\; , \;\;
\spa{a}.{\hat K_i} \neq 0 ,
\cr
{\cal J}_0 = & { \spa{\ell}.a \spa{\ell}.b \over \la \ell | K_1 K_2 | \ell \ra }
\longrightarrow J_0[a,b; \{K_j\}]=
  {  \la a | [ K_1,  K_2] | b \ra  \over  \Delta_3 }, 
\cr 
{\cal J}_1^0= & { \spa{\ell}.a \spa{\ell}.b \spa{\ell}.c \over \la \ell | K_1 K_2 | \ell \ra \spa{\ell}.d }
\longrightarrow 
\cr
J_1^0[d;a,b,c; \{K_j\}]= & {  \la b | [K_1, K_2]  | d \ra \la c | [K_1 ,K_2 ] | a \ra +\Delta_3 \spa{b}.d\spa{c}.a 
  \over  2\Delta_3 \la d | K_1 K_2  | d \ra }
- { \spa{d}.b \spa{d}.c   \la a | [ K_1 ,K_2] | d \ra
\over 2 \la d | K_1 K_2  | d \ra^2  }, 
\cr}
\equn
$$
where,
$$
\Delta_3=4(K_1\cdot K_2)^2-4K_1^2K_2^2 = (K_1^2)^2+  (K_2^2)^2+  (K_3^2)^2-2 (  K_1^2K_2^2+  K_2^2K_3^2 +K_3^2K_1^2).
\equn
$$
The above expressions are valid for $\ell=\ell_0$, $\ell=\ell_1$ and $\ell=\ell_2$ since,
$$
{\la a | [ K_1,K_3]|b\ra \over 2\la a | K_1K_3|a\ra  } ={\la a | [ K_1,K_2]|b\ra \over 2\la a | K_1K_2|a\ra  } 
={\la a | [ K_2,K_3]|b\ra \over 2\la a | K_2K_3|a\ra  }.
\equn 
$$

We can, as usual, extend these to, for example,
$$ 
{\cal J}_n^0=  { \spa{\ell}.a \spa{\ell}.b \prod_{i=1}^n\spa{\ell}.{c_i} \over \la\ell | K_1 K_3 | \ell \ra  \prod_{i=1}^n \spa{\ell}.{d_i} }
\longrightarrow J_n^0[\{ d_i\} ; \{a,b,c_i\} ; \{K_j\}]\, .
\equn
$$
The $J_n^0$ can be expressed in terms of $J_1^0$ as, 
$$
J_n^0[\{ d_i\} ; \{a,b,c_i\} ; \{K_j\}]
= \sum_{i=1}^n 
C_i
J_1^0[d_i; a,b,c_n; \{K_j\} ] ,
\equn
$$
with,
$$
C_i = {
\prod_{j=1}^{n-1} \spa{c_j}.{d_i} 
\over 
\prod_{j=1,j\neq i}^n \spa{d_j}.{d_i}
}\, .
\equn
$$

We also have terms which are of order $\ell^1$:
$$
\small
\eqalign{
&{\cal E}_0^1=[A|\ell_0|a\ra \longrightarrow  E_0^1[a;A;\{K_j\}]= -[A|a_0|a\ra \, , 
\cr
&{\cal E}_1^1={ [A|\ell_0|b\ra \spa{\ell_0}.c \over \spa{\ell_0}.d }
\longrightarrow 
\cr
& E_1^1[d;b,c;A;\{K_j\}]=-{  [A|a_0|b \ra\la d |[K_1,K_2]|c\ra + [A|a_0|c\ra\la d |[K_1,K_2]|b\ra - [A|a_0|d\ra\la b |[K_1,K_2]|c\ra
\over 2 \la d | K_1K_2|d\ra }
\cr
&\hskip 4.0truecm
+{ \Delta_3 [A|a_0|d\ra \spa{d}.b \spa{d}.c 
\over 2 \la d | K_1K_2|d\ra^2 } \, ,
\cr
&{\cal J}_0^1={ [A|\ell_0|b\ra \spa{\ell_0}.c  \spa{\ell_0}.d \over \la \ell_0 | K_1 K_2 | \ell_0 \ra }
\longrightarrow 
\cr
&
 J_0^1[b,c,d;A;\{K_j\}] =  { [A|a_0|b \ra\la c |[K_1,K_2]|d\ra 
+[A|a_0|c\ra \la d |[K_1,K_2]|b\ra+[A|a_0|d\ra \la b |[K_1,K_2]|c\ra ) \ra
\over 8 \Delta_3 }\, ,
\cr}
\equn
$$
where,
$$
a_0^\mu  = { -K_3^2 ( K_1^2+K_2^2-K_3^2 )\over \Delta_3  }   K_1^\mu 
+{ K_1^2 ( K_3^2+K_2^2-K_1^2 )\over \Delta_3  }  K_3^\mu.
\equn
$$
Expressions for $\ell=\ell_1$ are obtained by replacing $a_0$ by $a_1$ where,
$$
a_1^\mu= { -K_1^2 ( K_2^2+K_3^2-K_2^2 )\over \Delta_3  }   K_2^\mu 
+{ K_2^2 ( K_1^2+K_3^2-K_2^2 )\over \Delta_3  }  K_1^\mu
=a_0^\mu -K_1^\mu \, .
\equn$$
The $\ell=\ell_2$ expressions are obtained in a similar fashion.

Finally for order $\ell^1$ we have,
$$
\eqalign{
{\cal J}^1_1& = { [A|\ell_0|b\ra \spa{\ell_0}.c  \spa{\ell_0}.d  \spa{\ell_0}.e \over \la \ell_0 | K_1 K_2 | \ell_0 \ra  \spa{\ell_0}.f }
\longrightarrow  
\cr
J^1_1[f; b,c,d,e;A;\{K_j\} ]
&=
\sum_{P_{12}} {  [A|a_0|b \ra \la f |[K_1,K_2]|c \ra \la d |[K_1,K_2]|e \ra \over  4 \Delta_3\la f | K_1K_2|f\ra } 
\cr
&
-\sum_{P_6} {  [A|a_0|f \ra \la b |[K_1,K_2]|c \ra \la d |[K_1,K_2]|e \ra \over 24 \Delta_3\la f | K_1K_2|f\ra } 
\cr
&
-\sum_{P_6} {  [A|a_0|f \ra \spa{f}.b  \spa{f}.c \la d |[K_1,K_2]|e \ra
\over 12 \la f | K_1K_2|f\ra^2 } 
\cr
&
-\sum_{P_4}
{  [A|a_0|b \ra\la f |[K_1,K_2]|c \ra\la f |[K_1,K_2]|d \ra\la f |[K_1,K_2]|e \ra
\over 4 \Delta_3\la f | K_1K_2|f\ra^2 } 
\cr
&-\sum_{P_6} {  [A|a_0|f \ra \spa{f}.b  \spa{f}.c \la f |[K_1,K_2]|d \ra\la f |[K_1,K_2]|e \ra
\over 12 \la f | K_1K_2|f\ra^3 } .
 \cr}
\equn
$$

For QCD amplitude we generically need forms of order $\ell^2$, 
$$
\eqalign{
&{\cal E}^2_0 = [A|\ell_0|a\ra [B|\ell_0|b\ra \longrightarrow 
\cr
&
E^2_0[a,b;A,B; \{K_j\} ]  = -{1 \over 2} \sum_{P_2} [A|a_0|a\ra [B|a_0|b\ra 
-{ K_1^2K_2^2K_3^2\over 2 \Delta_3^2 }
[A|[K_1,K_2]|B]\la a |[K_1,K_2]|b\ra \, ,
\cr
& {\cal E}^2_1 = { [A|\ell_0|a\ra [B|\ell_0|b\ra \spa{\ell_0}.c \over \spa{\ell_0}.d } 
\longrightarrow
\cr
& E^2_1[d; a,b,c ;A,B; \{K_j\} ] =
-{   [A|a_0|d\ra[B|a_0|d\ra 
\la d |[K_1,K_2]|a\ra
\la d |[K_1,K_2]|b\ra
\la d |[K_1,K_2]|c\ra
\over 2 \la d | K_1K_2|d\ra^3 }
\cr
 & \hskip 2.5truecm 
+\sum_{P_6} {  [A|a_0|d\ra[B|a_0|a \ra \la d |[K_1,K_2]|b\ra \la d |[K_1,K_2]|c\ra
\over 3 \la d | K_1K_2|d\ra^2 }
\cr
&\hskip 2.5truecm 
+  [A|a_0|d\ra[B|a_0|d \ra \sum_{P_3 } {\la d |[K_1,K_2]|a\ra \la b |[K_1,K_2]|c \ra
\over 6 \la d | K_1K_2|d\ra^2 }
\cr
&  \hskip 2.5truecm 
-\sum_{P_6}
{ [A|a_0|d\ra[ B|a_0|a \ra \la b |[K_1,K_2]|c\ra 
\over  4 \la d | K_1K_2|d\ra  }
\cr
& \hskip 2.5truecm 
-\sum_{P_6} {[A|a_0|a\ra[ B|a_0|b \ra \la d |[K_1,K_2]|c\ra 
\over  3 \la d | K_1K_2|d\ra  }
\cr
&  \hskip 2.5truecm 
-\sum_{P_3} 
{ K_1^2K_2^2K_3^2  [ A   |[K_1,K_2]| B]  \la d |[K_1,K_2]|a\ra \la b |[K_1,K_2]|c\ra 
\over 12 \Delta_3^2 \la d | K_1K_2|d\ra  } \, ,
\cr
&
{\cal J}^2_0 ={ [A|\ell_0|a\ra [B|\ell_0|b\ra \spa{\ell_0}.c \spa{\ell_0}.d \over \la \ell_0 | K_1 K_2 | \ell_0 \ra  } 
\longrightarrow
\cr
 &J^2_0[a,b,c,d;A,B;\{K_j\}]=
{ 1 \over 6 \Delta_3 } \sum_{P_{12} } 
[A|a_0|a\ra[ B|a_0|b \ra \la c |[K_1,K_2]|d \ra 
\cr 
&  \hskip 2.5truecm \hskip 2.2truecm 
+{ K_1^2K_2^2K_3^2 \over 6 \Delta_3^3  }  \sum_{P_3} 
[ A |[K_1,K_2]|B ]\la a |[K_1,K_2]|b \ra\la c |[K_1,K_2]|d \ra \, ,
\cr}
$$
\vfill\eject

$$
\eqalign{
{\cal J}^2_1 = 
{ [A|\ell_0|a\ra [B|\ell_0|b\ra \spa{\ell_0}.c \spa{\ell_0}.d \spa{\ell_0}.e \over \la \ell_0 | K_1 K_2 | \ell_0 \ra  \spa{\ell_0}.f } 
\hskip -2truecm & \hskip +2truecm  \longrightarrow 
\cr J^2_1[f;a,b,c,d,e;A,B;\{K_j\}] = \hskip 0.3 truecm &
\cr 
{  K_1^2K_2^2 K_3^2 [ A |[K_1,K_2]|B ] \over 20  \la f | K_1K_2|f \ra \Delta_3^3 }    \sum_{P_{15}}&
\la f |[K_1,K_2]|a \ra\la b |[K_1,K_2]|c \ra
\la d |[K_1,K_2]|e \ra
\cr 
-{  1\over 60 \la f | K_1K_2|f \ra \Delta_3 }    \sum_{P_{30}}&
[A|a_0|f\ra[ B|a_0|a \ra  
\la b|[K_1,K_2]|c \ra\la d |[K_1,K_2]|e \ra
\cr 
+{  1\over 40 \la f | K_1K_2|f \ra \Delta_3 }    \sum_{P_{60}}&
[A|a_0|a\ra[ B|a_0|b \ra   \la c |[K_1,K_2]|d \ra\la f |[K_1,K_2]|e \ra
\cr 
-{ K_1^2K_2^2K_3^2 [ A |[K_1,K_2]|B ]  \over 40\la f | K_1K_2|f \ra^2 \Delta_3^3}   \sum_{P_{10}} &
\la a |[K_1,K_2]|b \ra\la f |[K_1,K_2]|c \ra
\la f |[K_1,K_2]|d \ra\la f |[K_1,K_2]|e \ra
\cr 
-{ 1\over 40 \la f | K_1K_2|f \ra^2 }   \sum_{P_{60}} &
[A|a_0|a\ra[ B|a_0|f \ra  
\la b|[K_1,K_2]|c \ra
\spa{f}.d \spa{f}.e 
\cr 
-{ 1 \over 30 \la f | K_1K_2|f \ra^2 }    \sum_{P_{60}} &
 [A|a_0|a\ra[ B|a_0| b\ra  
 \la f |[K_1,K_2]|c \ra
\spa{f}.d \spa{f}.e   
\cr
+{ 1 \over 30 \la f | K_1K_2|f \ra^3 }    \sum_{P_{60}} &
 [A|a_0|a\ra[ B|a_0|f \ra  
  \la f |[K_1,K_2]|b \ra\la f |[K_1,K_2]|c \ra
\spa{f}.d \spa{f}.e     \cr 
-{ [A|a_0|f\ra[ B|a_0|f \ra  \over 60 \la f | K_1K_2|f \ra^3 }    \sum_{P_{30}}& 
\la a |[K_1,K_2]|b \ra  \la f |[K_1,K_2]|c \ra
\spa{f}.d \spa{f}.e 
\cr 
-{ [A|a_0|f\ra[ B|a_0|f \ra  \over 20\la f | K_1K_2|f \ra^4 }    \sum_{P_{10}} &
\la f |[K_1,K_2]|a \ra
\la f |[K_1,K_2]|b \ra\la f |[K_1,K_2]|c \ra
\spa{f}.d \spa{f}.e  \, .
\cr}
\equn
$$

Cubic and higher order terms can also be evaluated. For example, 
$$
\eqalign{
[A|\ell_0|a\ra [B|\ell_0|b\ra [C|\ell_0|c\ra \longrightarrow &
-{1\over 6} \sum_{P_{6} }[A|a_0|a\ra [B|a_0|b\ra [C|a_0|c\ra
\cr
&
-{ K_1^2K_2^2K_3^2\over 6\Delta_3^2 }\sum_{P_{9}}
[A|[K_1,K_2]|B]\la a |[K_1,K_2]|b\ra [C|a_0|c\ra \, ,
\cr}
\equn
$$
however we can, in general, avoid these in  QCD calculations. 

\section{All-$n$ Expressions for Coefficients}
\label{BoxAppendix}

\subsection{Box Coefficients} 

Many of the box coefficients that appear in the seven-point $\NeqOne$ amplitudes are special 
cases of general $n$-point expressions. We gather these together here and give their specialisations to the 
box coefficients of section~\ref{SevenPointSection}. 
We denote the external legs contributing to the non-null momentum $K_i$ by $\{ f_i,\cdots, u_i \}$.

The three-mass boxes in a NMHV $\NeqOne$ amplitude vanish unless exactly one negative helicity gluon is attached to each non-null vertex. We denote this single
negative helicity gluon on vertex $i$ by $m_{i-1}$. 
\begin{center}\begin{picture}(100,120)(0,0)
\Line(25,25)(25,75)
\Line(25,25)(75,25)
\Line(25,75)(75,75)
\Line(75,25)(75,75)
\Line(95,5)(75,25)
\Text(95,0)[]{$s_1^+$}
\Line(75,75)(75,95)
\Line(75,75)(95,75)
\Line(75,75)(90,90)
\Text(70,103)[]{$f_4$}
\Text(100,67)[]{$u_4$}
\Text(95,95)[]{$m_3^-$}
\Text(95,115)[]{$\small{S_4}$}
\DashCArc(75,75)(15,0,90){1}
\Line(25,75)(25,95)
\Line(25,75)(5,75)
\Line(25,75)(10,90)
\Text(30,103)[]{$u_3$}
\Text(0,67)[]{$f_3$}
\Text(2,95)[]{$m_2^-$}
\Text(5,115)[]{$\small{S_3}$}
\DashCArc(25,75)(15,90,180){1}
\Line(25,25)(25,5)
\Line(25,25)(5,25)
\Line(25,25)(10,10)
\Text(25,-3)[]{$f_2$}
\Text(-3,25)[]{$u_2$}
\Text(2,5)[]{$m_1^-$}
\Text(0,40)[]{$\small{S_2}$}
\DashCArc(25,25)(15,180,270){1}
\end{picture}
\end{center}

$$
\eqalign{
 c^{3m}[m_1,m_2,m_3;& d,K_2,K_3,K_3]= -{
{\cal H}^2 \spa{u_2}.{f_3}\spa{u_3}.{f_4}\spa{m_3}.d\spa{m_1}.d\over 2 \spa{1}.{2}\spa{2}.{3}\ldots \spa{n}.{1} }
\times
\cr
&
 {
 \la m_1| K_3K_4|d\ra 
\la m_3 |K_3K_2|d\ra 
\la m_2 |K_3K_2|d\ra 
\la m_2 |K_3K_4|d\ra [d | K_2 K_3 K_4 | d\ra 
\over 
\la d| K_2 K_3 |d \ra^2
\BRDM{d}{K_4}{K_3}{u_2}\BRDM{d}{K_4}{K_3}{f_3}
\BRDM{d}{K_2}{K_3}{u_3}\BRDM{d}{K_2}{K_3}{f_4}
 K^2_3
}
\cr}
\equn
$$
where,
$$
{\cal H} =
\spa{m_1}.{m_2}\BRDM{d}{K_2}{K_3}{m_3}   +\spa{m_3}.{m_2}\BRDM{d}{K_4}{K_3}{m_1}. \, \, \, \,\,   
\equn
$$
These box coefficients are close in form to those of $\NeqFour$ Yang-Mills~\cite{Bern:2004jelly,Bidder:2005in}. 


All of the NMHV two mass boxes we need can be found using Generalised Unitarity taking MHV and $\overline{\rm MHV}$ 
tree amplitudes as the inputs. 
\begin{center}\begin{picture}(100,120)(0,0)
\Line(25,25)(25,75)
\Line(25,25)(75,25)
\Line(25,75)(75,75)
\Line(75,25)(75,75)
\Line(5,5)(25,25)
\Text(0,0)[]{$b^+$}
\Line(95,5)(75,25)
\Text(95,0)[]{$m_1^-$}
\Line(75,75)(75,95)
\Line(75,75)(95,75)
\Line(75,75)(90,90)
\Text(70,103)[]{$f_4$}
\Text(100,67)[]{$u_4$}
\Text(95,95)[]{$m_3^-$}
\Text(95,115)[]{$\small{K_4}$}
\DashCArc(75,75)(15,0,90){1}
\Line(25,75)(25,95)
\Line(25,75)(5,75)
\Line(25,75)(10,90)
\Text(30,103)[]{$u_3$}
\Text(0,67)[]{$f_3$}
\Text(2,95)[]{$m_2^-$}
\Text(5,115)[]{$\small{K_3}$}
\DashCArc(25,75)(15,90,180){1}
\end{picture}
\begin{picture}(150,170)(-50,0)
\Text(70,103)[]{$f_4$}
\Text(100,67)[]{$u_4$}
\Line(25,25)(25,75)
\Line(25,25)(75,25)
\Line(25,75)(75,75)
\Line(75,25)(75,75)
\Line(25,25)(25,5)
\Line(25,25)(5,25)
\Line(25,25)(10,10)
\Text(25,-3)[]{$f_2$}
\Text(-3,25)[]{$u_2$}
\Text(2,5)[]{$p^+$}
\Text(0,40)[]{$\small{K_2}$}
\DashCArc(25,25)(15,180,270){1}
\Line(95,5)(75,25)
\Text(95,-3)[]{$b^+$}
\Line(75,75)(75,95)
\Line(75,75)(95,75)
\Line(75,75)(90,90)
\Text(95,98)[]{$m^-$}
\Text(95,115)[]{$\small{K_4}$}
\DashCArc(75,75)(15,0,90){1}
\Line(25,75)(5,95)
\Text(2,101)[]{$a^+$}
\end{picture}\end{center}
With the
labellings given in the figure, the coefficients are,
$$
\eqalign{
c^{2mh}=&{P^2\spb{m_1}.{b}^2\la m_3|P|m_1]\la m_2|P|m_1]\spa{m_2}.b\la m_3|K_4P|b\ra\over
2 K_4^2[m_1|K_3P|m_1]\la b|PK_4|b\ra\la f_4|P|m_1]\la u_3|P|m_1]\la b|PK_4|u_4\ra\spa{f_3}.{b}[m_1|P|b\ra^2}
\cr
&\times{(P^2\la b|m_1P|b\ra\spa{m_2}.{m_3}+\la b|PK_4|b\ra\la m_3|m_1P|m_2\ra)^2
\over
\prod_{i=f_3}^{u_3-1} \spa{i}.{i+1}
\prod_{i=f_4}^{u_4-1} \spa{i}.{i+1}  },
\cr
c^{2me}= &
{[p|K_2|m\ra^2 [b|K_2|a\ra [a|K_2|b\ra [p|K_2|a\ra [p|K_2|b\ra \spa{m}.{a}\spa{m}.{b} 
\over
2 
\spa{a}.{f_4} \spa{u_4}.b \prod_{i=f_4}^{u_4-1} \spa{i}.{i+1} 
\prod_{i=f_2}^{u_2-1} \spb{i}.{i+1} 
\spa{a}.{b}^2[ f_2|K_2|a\ra[u_2|K_2|b\ra K_2^2} \, ,
\cr}
\equn
$$
where,
$P=K_3+k_b=-(K_4+k_{m_1})$. 

\def\redundantfigure{
\begin{center}\begin{picture}(100,120)(0,0)
\Line(25,25)(25,75)
\Line(25,25)(75,25)
\Line(25,75)(75,75)
\Line(75,25)(75,75)
\Text(17,50)[]{$\ell_1$}
\Text(50,17)[]{$\ell_4$}
\Text(50,83)[]{$\ell_2$}
\Text(83,50)[]{$\ell_3$}
\Line(25,25)(25,5)
\Line(25,25)(5,25)
\Line(25,25)(10,10)
\Text(25,0)[]{$b+^-$}
\Text(0,25)[]{$a-^-$}
\Text(2,5)[]{$p^+$}
\Text(0,37)[]{$\small{K_4}$}
\DashCArc(25,25)(15,180,270){1}
\Line(95,5)(75,25)
\Text(95,0)[]{$b^+$}
\Line(75,75)(75,95)
\Line(75,75)(95,75)
\Line(75,75)(90,90)
\Text(95,98)[]{$m^-$}
\Text(95,115)[]{$\small{K_2}$}
\DashCArc(75,75)(15,0,90){1}
\Line(25,75)(5,95)
\Text(2,98)[]{$a^+$}
\end{picture}\end{center}
}


One mass boxes in $\NeqOne$ amplitudes have either one or two negative helicity external legs attached to the 
three-point corners. In the latter case, the third negative helicity leg of any NMHV amplitude attaches to an MHV corner and a general form
exists:
\begin{center}\begin{picture}(100,120)(0,0)
\Line(25,25)(25,75)
\Line(25,25)(75,25)
\Line(25,75)(75,75)
\Line(75,25)(75,75)
\Line(5,5)(25,25)
\Text(0,0)[]{$b^+$}
\Line(95,5)(75,25)
\Text(95,0)[]{$m_1^-$}
\Line(75,75)(75,95)
\Line(75,75)(95,75)
\Line(75,75)(90,90)
\Text(73,103)[]{$f_4$}
\Text(100,70)[]{$u_4$}
\Text(95,98)[]{$m^-_3$}
\Text(95,115)[]{$\small{K_4}$}
\DashCArc(75,75)(15,0,90){1}
\Line(25,75)(5,95)
\Text(-3,100)[]{$m_2^-$}
\end{picture}\end{center}

$$
c^{1m}_0=-{[b|K_4|m_3\ra^2[m_1|K_4|m_3\ra[m_2|K_1|m_3\ra s_{m_1b}s_{bm_2}\over
2 K_4^2\spb{m_1}.{m_2}^2[m_1|K_4|m_2+1\ra[m_2|K_4|u_4\ra\prod_{i=f_4}^{u_4-1} \spa{i}.{i+1} }.
\equn
$$
When there are two negative helicity external legs attached to the massive corner, NMHV tree amplitudes
are required and we must consider each helicity configuration separately:

\FIGURE{
\begin{minipage}{0.23\linewidth}
\centering
\begin{center}\begin{picture}(80,105)(0,0)
\Line(20,20)(20,60)
\Line(20,20)(60,20)
\Line(20,60)(60,60)
\Line(60,20)(60,60)
\Line(5,5)(20,20)
\Text(0,0)[]{$b^-$}
\Line(75,5)(60,20)
\Text(75,0)[]{$a^+$}
\Line(60,60)(60,77)
\Line(60,60)(77,60)
\Line(60,60)(76,68)
\Line(60,60)(68,76)
\Text(60,85)[]{$d^+$}
\Text(86,58)[]{$g^-$}
\Text(75,83)[]{$e^+$}
\Text(86,72)[]{$f^-$}
\Text(40,100)[]{$\small{1}$}
\Line(20,60)(5,75)
\Text(5,83)[]{$c^+$}
\end{picture}\end{center}
\end{minipage}
\begin{minipage}{0.23\linewidth}
\centering
\begin{center}\begin{picture}(80,105)(0,0)
\Line(20,20)(20,60)
\Line(20,20)(60,20)
\Line(20,60)(60,60)
\Line(60,20)(60,60)
\Line(5,5)(20,20)
\Text(0,0)[]{$b^-$}
\Line(75,5)(60,20)
\Text(75,0)[]{$a^+$}
\Line(60,60)(60,77)
\Line(60,60)(77,60)
\Line(60,60)(76,68)
\Line(60,60)(68,76)
\Text(60,85)[]{$d^-$}
\Text(86,58)[]{$g^-$}
\Text(75,83)[]{$e^+$}
\Text(86,72)[]{$f^+$}
\Text(40,100)[]{$\small{2}$}
\Line(20,60)(5,75)
\Text(5,83)[]{$c^+$}
\end{picture}\end{center}
\end{minipage}
\begin{minipage}{0.23\linewidth}
\centering
\begin{center}\begin{picture}(80,105)(0,0)
\Line(20,20)(20,60)
\Line(20,20)(60,20)
\Line(20,60)(60,60)
\Line(60,20)(60,60)
\Line(5,5)(20,20)
\Text(0,0)[]{$b^-$}
\Line(75,5)(60,20)
\Text(75,0)[]{$a^+$}
\Line(60,60)(60,77)
\Line(60,60)(77,60)
\Line(60,60)(76,68)
\Line(60,60)(68,76)
\Text(60,85)[]{$d^+$}
\Text(86,58)[]{$g^-$}
\Text(75,83)[]{$e^-$}
\Text(86,72)[]{$f^+$}
\Text(40,100)[]{$\small{3}$}
\Line(20,60)(5,75)
\Text(5,83)[]{$c^+$}
\end{picture}\end{center}
\end{minipage}
\begin{minipage}{0.23\linewidth}
\centering
\begin{center}\begin{picture}(80,105)(0,0)
\Line(20,20)(20,60)
\Line(20,20)(60,20)
\Line(20,60)(60,60)
\Line(60,20)(60,60)
\Line(5,5)(20,20)
\Text(0,0)[]{$b^-$}
\Line(75,5)(60,20)
\Text(75,0)[]{$a^+$}
\Line(60,60)(60,77)
\Line(60,60)(77,60)
\Line(60,60)(76,68)
\Line(60,60)(68,76)
\Text(60,85)[]{$d^+$}
\Text(86,58)[]{$g^+$}
\Text(75,83)[]{$e^-$}
\Text(86,72)[]{$f^-$}
\Text(40,100)[]{$\small{4}$}
\Line(20,60)(5,75)
\Text(5,83)[]{$c^+$}
\end{picture}\end{center}
\end{minipage}
}

$$
\eqalign{
c_1^{1m}&={\la b|P_{abc}P_{def}|f\ra^2\la c |P_{abc}P_{def}|f\ra s_{ab}s_{bc}\over 2 t_{def}t_{abc}\spa{a}.{c}^2\spa{d}.e\spa{e}.f[g|P_{ab}|c\ra[g|P_{def}|d\ra}
-{[e|P_{efg}|b\ra^2[e|P_{efg}|c\ra s_{ab}s_{bc}\over 2\spa{a}.c^2\spb{e}.f\spb{f}.g\spa{c}.d[g|P_{efg}|d\ra t_{efg}} \, ,
\cr
\cr
c_2^{1m}&={s_{ab}s_{bc}\spa{b}.g^2\spa{c}.g\spb{e}.f^3\over 2 t_{def}\spa{a}.c^2\spb{d}.e[d|P_{def}|g\ra[f|P_{def}|c\ra}
\cr
+& {s_{ab}s_{bc}(\spb{f}.g\spa{g}.d\spa{b}.c-[f|P_{abc}|c\ra\spa{d}.b)^2(\spb{f}.g\spa{g}.d\spa{a}.c-[f|P_{abc}|c\ra\spa{d}.a)[f|P_{abc}|c\ra\over 2\spa{a}.c^2\spb{f}.g[g|P_{abc}|c\ra\spa{d}.e[f|P_{de}|c\ra\la c|P_{abc}P_{fg}|e\ra(s_{de}\spa{c}.a+\la c|P_{de}P_{bc}|a\ra)} 
\cr
&
-{s_{ab}s_{bc}\la b|P_{abc}P_{efg}|g\ra^2\la a|P_{abc}P_{efg}|g\ra\la c|P_{abc}P_{efg}|g\ra\over 2 t_{efg}t_{abc}\spa{a}.c^2[d|P_{efg}|g\ra[d|P_{abc}|a\ra\spa{e}.f\spa{f}.g\la c|P_{abc}P_{efg}|e\ra}
\, ,
\cr} 
$$
$$
\eqalign{
c_3^{1m}&={s_{ab}s_{bc}\la b|P_{abc}P_{def}|e\ra^2\la c|P_{abc}P_{def}|e\ra\la a|P_{abc}P_{def}|e\ra\over 2 t_{def}t_{abc}\spa{c}.a^2\spa{d}.e\spa{e}.f[g|P_{abc}|c\ra[g|P_{def}|d\ra\la a|P_{abc}P_{def}|f\ra}
\cr
&-{s_{ab}s_{bc}[f|P_{efg}|b\ra^2[f|P_{efg}|a\ra[f|P_{efg}|c\ra\over 2 t_{efg}\spa{a}.c^2\spa{c}.d\spb{e}.f\spb{f}.g[e|P_{efg}|a\ra[g|P_{efg}|d\ra}
\cr
&-{s_{ab}s_{bc}([d|P_{abc}|a\ra\spa{g}.b+\spb{d}.e\spa{e}.g\spa{a}.b)^2([d|P_{abc}|a\ra\spa{g}.c+\spb{d}.e\spa{e}.g\spa{a}.c)\over 2
\spa{c}.a^2\spa{f}.g\spb{d}.e[e|P_{efg}|a\ra\la|P_{abc}P_{def}|f\ra(s_{de}\spa{c}.a+\la c|P_{de}P_{abc}|a\ra)}  \, ,
\cr
\cr
c_4^{1m}&=
-{s_{ab}s_{bc}\spa{f}.a(\spa{f}.a[d|P_{abc}|b\ra-\spa{a}.b\spb{d}.g\spa{g}.f)^2(\spa{f}.a[d|P_{abc}|c\ra-\spa{a}.c\spb{d}.{g}\spa{g}.f)\over 2\spa{c}.a^2\spa{f}.g\spa{g}.a\spb{d}.e\la a|P_{abc}P_{def}|f\ra[e|P_{efg}|a\ra(s_{de}\spa{c}.a+\la c|P_{de}P_{abc}|a\ra)}  
\cr
&+
{s_{ab}s_{bc}[g|P_{abc}|b\ra^2[g|P_{abc}|a\ra\spa{e}.f^3\over 2\spa{a}.c^2t_{def}t_{abc}\spa{d}.e[g|P_{def}|d\ra\la a|P_{abc}P_{def}|f\ra}
\cr
&-{s_{ab}s_{bc}[g|P_{efg}|b\ra^2[g|P_{efg}|a\ra[g|P_{efg}|c\ra\over 2 t_{efg}\spa{a}.c^2\spa{c}.d\spb{e}.f\spb{f}.g[e|P_{efg}|a\ra[g|P_{efg}|d\ra}
\; . 
\cr}
\equn
$$

\subsection{Particular Bubble Coefficients} 

Our first example is the bubble coefficient in an $\NeqOne$ MHV amplitude. Coefficients of $\ln(-P^2_{a\cdots b})$
vanish unless exactly one of the two negative helicity legs lies within $P_{a\cdots b}$.
The non-vanishing coefficient of $\ln(-P^2_{a\cdots b})$
is then,
$$
{ \spa{m_1}.{m_2}^2 \spa{b}.{b+1} \spa{a-1}.{a} \over \prod_i \spa{i}.{i+1} } 
\times H_4 [ a-1,a,b,b+1 ; m_1, m_1 , m_2 , m_2 ; P_{a\cdots b}] \, .
\equn
$$

The second example is relevant for the seven-point NMHV amplitudes. Consider the case where one side of the cut is MHV and the other is $\overline{\rm MHV}$. 
Let $m$ be the single negative helicity on the MHV side which contains legs $a\cdots b$ and 
$p$ be the single positive helicity on the  $\overline{\rm MHV}$ side. 
The  coefficient of $\ln(-P^2_{a\cdots b})$
is then just,
$$
\eqalign{
{ [p|P_{a\cdots b}|m\ra^2 \over \prod_{i=a}^{b-1}  \spa{i}.{i+1} \prod_{j=b+1}^{a-2}  \spb{j}.{j+1} P_{a\cdots b}^2  } 
\times
\cr
 H_4 [ a,b, P_{a\cdots b}|a-1]  ,  P_{a\cdots b}|b+1] ;  m, m,  P_{a\cdots b}|p], P_{a\cdots b}|p]  ; P_{a\cdots b}]  \, .
\cr}
\equn\label{N1MHVMHVbar}
$$ 
Since many of the $d_i$ coefficients in the seven-point NMHV amplitudes are 
special cases of this generic expression specialised to seven-point we define,
$$
\eqalign{
C_0[& a,b,c,d,e,f,g;p,m]\equiv
\cr
&{ [p|P_{gab}|m\ra^2 \over \spa{c}.d\spa{d}.e\spa{e}.f  \spb{g}.a\spb{a}.b  P_{gab}^2  } 
\times H_4 [c,f, P_{gab}|b]  ,  P_{gab}|g] ;  m, m,  P_{gab}|p], P_{gab}|p]  ; P_{a\cdots b}] \, .
\cr}
\equn\label{Mfunctiondef}
$$

\section{Seven-point Tree Expressions} 
\label{SevenPointTree}

We have:
$$A:A(s_1,\bar{s}_2 , 3^-,4^-, 5^+,6^+,7^+  ) 
= T^A_{1a}+T^A_{1b}+T^A_2+T^A_3 \, , 
\equn
$$
with,
$$
\eqalign{
T^A_{1a}= & {[5|P_{345}|2\ra[5|P_{345}|1\ra^2 
\over \spa6.7 \spa7.1\spa1.2\spb3.4 \spb4.5  [3|P_{345}|6\ra t_{345}}  
\times
\Bigl(
{ [5|P_{345}| 1\ra  \over  [5|P_{345}|2\ra }\Bigr)^{2h}  \, ,
\cr
T^A_{1b}= & 
{\la 6 | P_{71}  P_{23}|4\ra
( [2|P_{67}|1\ra \spa6.4 - [2|5|4\ra\spa1.6 )^2
\over\spa4.5 \spa5.6\spa6.7\spa7.1 \spb2.3 
[3|P_{712}|6 \ra 
[2|P_{71}|6 \ra
\la 1 | P_{23}P_{45}|6\ra  }  
\cr
&\hskip 6.3 truecm 
\times
\Bigl(
{ ( [2|P_{67}|1\ra \spa6.4 - [2|5|4\ra\spa1.6 ) \over\la 6 | P_{71}  P_{23}|4\ra  }
\Bigr)^{2h}  \, ,
\cr
T^A_2= & { \spa3.4^4 \over \spa3.5^4} \times T^B_2   \, ,
\cr
T^A_3=&  { [7|P_{456}|4\ra^4 \over [7| P_{456}| 5\ra^4 } \times T^B_3  \, .
\cr}
\equn
$$

$$B:A(s_1,\bar{s}_2 , 3^-,4^+, 5^-,6^+,7^+  ) 
= T^B_{1a}+T^B_{1b}+T^B_{1c}+T^B_2+T^B_3+T^B_4 \, ,
\equn
$$
with,
$$
\eqalign{
T^B_{1a}= & 
{ [4|P_{234}|5\ra^2 \spb2.4^2 \spa5.1^2   
\over \spa5.{6}\spa6.7\spa7.1 \spb2.3\spb3.4 t_{234}[2|P_{234}|5\ra [4|P_{234}|1\ra }
\times
\Bigl(
{\spb2.4 \spa5.1    \over [4|P_{234}|5\ra  
}
\Bigr)^{2h}  \, ,
\cr
T^B_{1b}= & 
{-\spa2.3\spa1.3^2 [4|P_{67}|1\ra^4 
\over \spa6.7\spa7.1\spa1.2 \spb4.5 
 [5|P_{67}|1\ra[4|P_{23}|1\ra
 \la 1 | P_{67} P_{45}|3\ra 
    \la 6 | P_{45}P_{23}|1\ra  }  
\times
\Bigl(   {  \spa1.3 \over \spa2.3} \Bigr)^{2h}  \, ,
\cr
T^B_{1c}= & 
{  t_{671}[2|P_{671} |1\ra^2 \spa3.5^4 \over 
 \spa6.7\spa7.1 \spa3.4\spa4.5 [2|P_{71}|6\ra [2|P_{345}|5\ra \la 1 |  P_{67} P_{45}|3\ra t_{345}  }
\times \Bigl(
{ -[2|P_{671}| 1\ra  \over t_{671}   } \Bigr)^{2h}   \, ,
\cr
T^B_{2}= &
 {  - \spb2.7^2\spb1.7  \spa3.5^4 \over 
\spb1.2 \spa3.4\spa4.5\spa5.{6}
[2 | P_{712}|6 \ra [7 | P_{712}|3 \ra  t_{712} }
\times \Bigl( -{ \spb2.7 \over  \spb1.7 } \Bigr)^{2h}  \, ,
\cr
T^B_{3}= & 
{ \spa1.3^2\spa2.3^2[7|P_{123}|5\ra^4 
\over \spa{1}.2\spa2.3 \spa4.5\spa5.{6}
[7|P_{123}|3\ra [7|P_{123}|4\ra
 \la 1 | P_{23}P_{45}|6\ra t_{123}t_{456} }
\times \Bigl( { \spa1.3 \over \spa2.3 } \Bigr)^{2h}  \, ,
\cr
T^B_{4}= & 
{- \spa1.3^2\spa2.3^2\spa5.6^2\spb6.7^3
\over
\spa{1}.2\spa2.3\spa3.4 \spa5.{6}
[7|P_{56}|4\ra  
[5 |  P_{567} |1\ra t_{567} s_{56} 
}\times \Bigl( { \spa1.3 \over \spa2.3 } \Bigr)^{2h}  \, .
\cr}
\equn
$$

$$
C: A(s_1,\bar{s}_2 , 3^-,4^+, 5^+,6^-,7^+  )
=T^C_{1a}+T^C_{1b}+T^C_{1c}+T^C_2+T^C_3+T^C_4  \, ,
\equn
$$
with,
$$
\eqalign{
T^C_{1a}=& 
{ \spb2.4^2 [4|P_{234}|6\ra^2 \spa1.6^2
\over 
\spb2.3\spb3.4 \spa5.{6}\spa6.7\spa7.1  [4|P_{234} | 1 \ra  [2|P_{234}|5\ra t_{234} }   
\times \Bigl({ -\spb2.4 \spa1.6  \over  [4|P_{234}|6\ra }
\Bigr)^{2h}  \, , 
\cr
T^C_{1b}=& -{ \spa1.3^2 \spa2.3 \spb4.5^3  \spa1.6^4  \over  
\spa6.7\spa7.1\spa1.2
[4|P_{234}|1\ra  
[5|P_{234}|1\ra  
\la 1 | P_{67} P_{45}|3\ra 
\la 1 | P_{23} P_{45}|6\ra   }
\Bigl( { \spa1.3 \over \spa2.3  }
\Bigr)^{2h}  \, , 
\cr
T^C_{1c}=& 
 {  [2|P_{345}|3\ra^2
\la 6 | P_{71}P_{345} | 3 \ra^2 \spa1.6^2
\over \spa6.7\spa{7}.{1} \spa3.4\spa4.5
[ 2 | P_{71}| 6 \ra 
[2|P_{345}|5\ra 
\la 1 | P_{671}P_{345}|3\ra t_{345}t_{671} }
\times
\Bigl( {[2|P_{345}|3\ra \spa1.6\over\la 6 | P_{671}P_{345}|3\ra }
\Bigr)^{2h}  \, ,
\cr
T^C_{2}=& 
{\spb1.7^2\spb2.7^2\spa3.6^4\over\spb7.1\spb1.2\spa3.4\spa4.5\spa5.6[7|P_{712}|3\ra [2|P_{712}|6\ra t_{712} }
\left(-{\spb2.7\over\spb1.7}\right)^{2h}  \, ,
\cr
T^C_{3}=& 
{\spa1.3^2\spa2.3^2[7|P_{456}|6\ra^4 \over \spa1.2\spa2.3\spa4.5\spa5.6\la1|P_{123}P_{456}|6\ra [7|P_{456}|4\ra
[7|P_{456}|3\ra\t{123}\t{456}}
\left({\spa1.3\over\spa2.3}\right)^{2h}  \, ,
\cr
T^C_{4}=& 
{\spb7.5^4\spa1.3^2\spa2.3^2 \over
\spa1.2\spa2.3\spa3.4\spb5.6\spb6.7
[5|P_{567}|1\ra[7|P_{567}|4\ra\t{567}
}\left({\spa3.1\over\spa3.2}\right)^{2h}  \, .
\cr}
\equn
$$

$$ D: A_7(s_1,\bar{s}_2, 3^-,4^+,5^+,6^+,7^-)=T^D_{1}+T^D_{2}+T^D_{4a}+T^D_{4b}+T^D_{4c} \, , 
\equn
$$
with,
$$
\eqalign{
T^D_{1}=& -
{ [6 | P_{71} | 3\ra^2 \spa3.2  \spb1.6^2 \over  
\spb6.7 \spb7.1   \spa3.4\spa4.5 [1 |  P_{671}  | 5 \ra [6 |  P_{671}  | 2 \ra t_{671}}  
\times \Bigl( { [6|P_{71}|3\ra  \over \spb1.6 \spa3.2 }
\Bigr)^{2h}\, ,
\cr
T^D_{2}=& -
{\spa{7}.1 \spa{7}.2^2[6|P_{712}|3\ra^3   \over  \spa1.2\spa3.4\spa4.5
[6|P_{712}|2\ra 
\la 5 | P_{34}  P_{12} | 7\ra t_{345} t_{712} }
\times
\Bigl(
{ \spa{7}.1 \over \spa{7}.2 } \Bigr)^{2h} \, ,
\cr
T^D_{4a}=& 
{ [4|P_{56}|7\ra^3 \spa{7}.2^2\spa{7}.1^2 \over
\spa5.6  \spa6.7\spa{7}.1\spa1.2\spb3.4
 [3|P_{12}|7\ra 
\la 7 | P_{56} P_{34}|2\ra
  \la 7 | P_{12}P_{34}|5\ra }
\times
\Bigl( { \spa{7}.1 \over \spa{7}.2 } \Bigr)^{2h} \, ,
\cr
T^D_{4b}=& 
{  \spa2.3 [1  | P_{56}| 7\ra^2 \la 7 | P_{56} P_{234}|3\ra^2  
\over
\spa5.6 \spa6.7 \spa3.4
[1|P_{567} | 5\ra  
[1|P_{234} | 4\ra 
\la 7 | P_{56} P_{234}|2\ra t_{234} t_{567} }
\times
\Bigl({ -\la 7 | P_{56} P_{234}|3\ra \over \spa2.3[1 | P_{56} | 7\ra  } \Bigr)^{2h} \, ,
\cr
T^D_{4c}=& 
{ [1|P_{123}|7\ra^2 [2 |P_{123}|7\ra^2 
\over
\spa4.5\spa5.6\spa6.7
\spb1.2\spb2.3 [ 3|P_{123}|7\ra [1|P_{123}|4\ra t_{123} }
\times
\Bigl({ -[2 |P_{123}|7\ra \over [1 |P_{123}|7\ra }\Bigr)^{2h} \, .
\cr}
\equn
$$

$$
E: A_7(s_1,\bar{s}_2, 3^+,4^-,5^-,6^+,7^+)
=T^E_{1a}+T^E_{1b}+T^E_{1c}+T^E_2+T^E_3+T^E_4 \, ,
\equn
$$
with,
$$
\eqalign{
T^E_{1a} = & 
{  \spa2.4^2 t_{671}    \la 1 | P_{67} P_{23}|4\ra^2 
\over 
\spa6.7\spa7.1\spa2.3\spa3.4   [5| P_{671} |1 \ra 
[5|P_{671}|2\ra
\la 6 | P_{71} P_{23}|4\ra t_{234} 
}
\Bigl( 
{- \la 1 |P_{67}P_{23}|4\ra  
\over  
\spa2.4 t_{671}   
}
\Bigr)^{2h}  \, ,
\cr
T^E_{1b} = & 
 { -[3|P_{71}|6\ra^2 \spb2.3\spa4.5^3 \spa1.6^2\over
\spa5.{6} \spa6.7 \spa7.1   [2|P_{71}|6\ra  [3|P_{45}|6\ra 
\la 6 | P_{71}P_{23}|4\ra  \la 1 |P_{23}P_{45}|6\ra  }
\Bigl({
-\spb2.3 \spa6.1  \over [3  | P_{71} | 6 \ra  
}\Bigr)^{2h}  \, ,
\cr
T^E_{1c} = & 
{ [3|P_{345} |2\ra^2  [3|P_{345} |1\ra^2 
\over \spa6.7\spa7.1 \spa1.2 \spb3.4\spb4.5  [5|P_{345}|2 \ra [3|P_{345}|6\ra t_{345} }
\Bigl({[3|P_{345}| 1 \ra  
\over[3|P_{345} |2\ra 
}\Bigr)^{2h}  \, ,
\cr
T^E_{2} = & \Bigl({ \spa4.5 \over \spa3.5} \Bigr)^4 T_2^{B} \, ,
\cr
T^E_{3} = & {  [7|P_{456}|1\ra^2   [7|P_{456}|2\ra^2  \spa4.5^3 
\over \spa1.2\spa2.{3}\spa5.{6} [7|P_{456}|3\ra   [7|P_{56}|4\ra
\la 1 | P_{23} P_{45} | 6 \ra t_{123}  t_{456} }
\times\Bigr( { [7|P_{456}|1\ra  \over [7|P_{456}|2\ra  } 
\Bigl)^{2h} \, ,
\cr
T^E_{4} = & \Bigl({ \spa1.4^2\spa2.4^2 \over \spa1.3^2\spa2.3^2 } \Bigr)
\Bigl({\spa2.3\spa1.4\over \spa1.3\spa2.4 }\Bigr)^{2h}
 T_4^{B} \, .
\cr}
\equn
$$

$$
F:A_7(s_1,\bar{s}_2, 3^+,4^-,5^+,6^-,7^+)
=T^F_{1a}+T^F_{1b}+T^F_{1c}+T^F_2+T^F_3+T^F_4 \, ,
\equn
$$
with,
$$
\eqalign{
T^F_{1a} = & {  [5|P_{234}|4\ra^2 \spa2.4^2[5|P_{71}|6\ra^2 \spa1.6^2
\over
\spa6.7\spa7.1\spa2.3\spa3.4[ 5 | P_{234}| 1\ra 
[5|P_{234}|2\ra
\la 6 | P_{71} P_{23}|4\ra t_{671} t_{234} }
\Bigl( {[5|P_{234}|4\ra  \spa1.6  \over  \spa2.4[ 5 | P_{234} | 6\ra   }
\Bigr)^{2h} \, ,
\cr
T^F_{1b} = & 
{ -\spb2.3^2[3|P_{71}|6\ra^2  \spa4.{6}^4 \spa1.6^2 \over
\spa6.7\spa7.1[2|P_{71}|6\ra \spb2.3\spa4.5\spa5.{6} 
[ 3|P_{45}|6 \ra \la 6 | P_{71} P_{23}|4\ra \la 1 | P_{23}P_{45}|6\ra  }
\Bigl( { -\spb2.3  \spa6.1 \over [3|P_{71}|6\ra    } 
\Bigr)^{2h} \, ,
\cr
T^F_{1c} = & 
{ \spa{6}.2^2 \spa1.6^2  \spb3.5^4 \over
\spa6.7\spa7.1\spa1.2 \spb3.4\spb4.5  [ 5|P_{345}|2\ra [3|P_{345}|6\ra t_{345} }
\Bigl( {  \spa6.1\over  \spa{6}.2 } \Bigr)^{2h} \, ,
\cr
T^F_{2} = & { \spa4.{6}^4 \over \spa3.{6}^4 } T_2^C \, ,
\cr
T^F_{3} = & {  \spa4.{6}^4 [7|P_{123}|1\ra^2  [7|P_{123}|2\ra^2  \over 
\spa1.2\spa2.3\spa4.5\spa5.{6}
[7|P_{123}|3\ra   [7|P_{123}|4\ra
\la 1 | P_{23} P_{45} | 6\ra  
t_{123}t_{456}  }
\times
\Bigl( { [7|P_{123}|1\ra\over [7|P_{123}|2\ra } \Bigr)^{2h}  \, ,
\cr
T^F_{4} = &  {  \spa1.4^{2+2h}\spa2.4^{2-2h}  \over \spa1.3^{2+2h}\spa2.3^{2-2h} }T_{4}^C \, .
\cr}
\equn
$$

\section{$D_X$ Functions} 
\label{DXap}

\noindent
The function $D_A$ was given in section~\ref{DAbody}. The remaining five $D_X$ functions are given by: 

\def\spba#1.#2.#3{[#1|P_{#2}|#3\ra}
\def\spab#1.#2.#3{\la#1|P_{#2}|#3]}
\def\spaba#1.#2.#3.#4{\la#1|P_{#2}P_{#3}|#4\ra}

\def\FAforbodyofpaper{
$$
\eqalign{
&\hskip-1.0truecm D_A(a,b,c,d,e,f,g)= 
\cr &\hskip-1.0truecm
-{ \spba {e}.{{cde}}.{a}^2  
\over  
\spa{a}.{b}\spa{f}.{g}\spb{c}.{d}\spb{d}.{e} \spba{c}.{{cde}}.{f}t_{cde}}
H_2[  b,   g ; a , P_{cde}|e];P_{ab}]
\cr &\hskip-1.0truecm
+
{\spa{f}.{d}  
\over  
\spb{a}.{b}\spa{d}.{e}\spa{e}.{f}\spa{f}.{g}\spba{c}.{{de}}.{f}}
\overline{H_4}[a,c,P_{cde}|f\ra, P_{ab} P_{abc}P_{de}|f\ra ;
     b,g,XA,XA;
     P_{ab}]
\cr &\hskip-1.0truecm
+
{1  
\over  
\spb{a}.{b}\spa{d}.{e}\spa{e}.{f}\spa{f}.{g}\spba{c}.{{de}}.{f}}
\overline{H_5}[a,P_{ab}|g\ra,c,P_{cde}|f\ra,P_{ab} P_{abc}P_{de}|f\ra ;
     b , P_{ab}|f\ra , P_{ef}|d\ra, XA, XA;
     P_{ab}]
\cr &\hskip-1.0truecm
+
{\spa{c}.{d}^4\spb{g}.{b}^2  
\over  
\spb{a}.{b}\spa{c}.{d}\spa{d}.{e}\spa{e}.{f}\spba{g }.{ {ab}}.{c}t_{gab}  }  
H_2[ b , P_{ab} P_{gab}|f\ra ;  a , P_{ab}|g], P_{ab}]   
\cr &\hskip-1.0truecm
+
{\spa{a}.{c}^2\spba{g}.{{abc}}.{d}^4   
\over  
\spa{a}.{b}\spa{d}.{e}\spa{e}.{f}\spba{g}.{{ab}}.{c}\spba{g}.{{abc}}.{d}t_{abc}t_{def}}  
H_2[ b, P_{abc}P_{de}|f\ra ;  a ,c, P_{ab}],
\cr} 
$$
where,
$$
\eqalign{
\bar\lambda_{XA}(a,b,c,d,e,f,g)=  &\bar\lambda_g\spa{g}.{a}\spb{a}.{b}\spa{f}.{d}
                                  +\bar\lambda_f\spa{f}.{a}\spb{a}.{b}\spa{f}.{d}
                                  +\bar\lambda_e\spa{e}.{d}\spb{a}.{b}\spa{f}.{a}\cr&
                                  +\bar\lambda_b\spa{f}.{g}\spb{g}.{c}\spa{c}.{d}
                                  +\bar\lambda_b\spba{c}.{abc}.{f}\spa{c}.{d}
                                  +\bar\lambda_b\spa{f}.{d}s_{ab}.              
\cr}
$$
}
\bigskip
$$
\small 
\eqalign{
 & \hskip-1.0truecm D_B[a,b,c,d,e,f,g]= \cr &\hskip-1.0truecm-{1  \over  \spa{e}.{f}\spa{f}.{g}\spa{a}.{b}\spb{c}.{d} }G_5[a,g, P_{ab} P_{cd}|e\ra ,P_{ab}|c], P_{efg}|d] ;b, e, e, X_{B1a}, X_{B1a},  P_{ab}|d] ;d; P_{efg} ;P_{ab} ]\cr &\hskip-1.0truecm+{\spa{b}.{c}^2  \over  \spa{a}.{b}\spa{f}.{g}\spb{d}.{e}}H_6[a, g,  P_{fg}|e] , P_{efg}|d] ,P_{fg}P_{de}|c\ra ,P_{defg}P_{de}|f\ra ;b , c,  P_{fg}|d], P_{fg}|d], P_{fg}|d], P_{fg}|d]; P_{ab} ]\cr &\hskip-1.0truecm-{\spa{c}.{e}^4 \spb{a}.{f}  \over  \spa{a}.{b}\spa{c}.{d}\spa{d}.{e}\spa{f}.{g}t_{cde} } H_5[ a, g, P_{ab}P_{cde}|f\ra , P_{ab} P_{cde}|e\ra , P_{fg} P_{cde}|c\ra  ; b, a, f , P_{fg}P_{cde}|b\ra , P_{fg} P_{cde}|b\ra ; P_{ab}]\cr &\hskip-1.0truecm-{\spa{c}.{e}^4 \spb{b}.{f} \over  \spa{a}.{b}\spa{c}.{d}\spa{d}.{e}\spa{f}.{g}t_{cde} }H_5[ a, g, P_{ab} P_{cde}|f\ra, P_{ab} P_{cde}|e\ra , P_{fg} P_{cde}|c\ra ; b, b, f , P_{fg} P_{cde}|b\ra, P_{fg} P_{cde}|b\ra ; P_{ab}]\cr &\hskip-1.0truecm-{\spa{c}.{e}^4  \spb{a}.{g}\over  \spa{a}.{b}\spa{c}.{d}\spa{d}.{e}\spa{f}.{g}t_{cde}} H_5[ a, g,  P_{ab} P_{cde}|f\ra, P_{ab} P_{cde}|e\ra , P_{fg} P_{cde}|c\ra ; b, a, g , P_{fg} P_{cde}|b\ra, P_{fg} P_{cde}|b\ra ; P_{ab}]\cr &\hskip-1.0truecm-{\spa{c}.{e}^4  \spb{b}.{g}\over  \spa{a}.{b}\spa{c}.{d}\spa{d}.{e}\spa{f}.{g}t_{cde}} H_5[ a, g,   P_{ab} P_{cde}|f\ra, P_{ab} P_{cde}|e\ra , P_{fg} P_{cde}|c\ra ; b, b, g ,  P_{fg} P_{cde}|b\ra, P_{fg} P_{cde}|b\ra ; P_{ab}]\cr &\hskip-1.0truecm+{\spa{c}.{e}^4\spb{g}.{a}^2\spa{a}.{b}^2\over  \spa{c}.{d}\spa{d}.{e}\spa{e}.{f}\spa{a}.{b}s_{ab}\spba{g }.{{gab}}.{c}t_{gab} }
H_2[ a ,  P_{ab} P_{gab}|f\ra ; b ,P_{ab}|g] ;P_{ab} ]
\cr &\hskip-1.0truecm-{\spba{ g}.{{abc}}.{e}^4\spa{b}.{c}^2  \over   \spa{a}.{b}\spa{d}.{e}\spa{e}.{f}\spba{g}.{{abc}}.{c}\spba{g}.{{abc}}.{d}t_{abc}t_{def} }
H_2[ a, P_{abc} P_{de}|f\ra ;  b ,c; P_{ab}]
\cr &\hskip-1.0truecm-{\spb{f}.{g}^3\spa{b}.{c}^2  \over  \spa{a}.{b}\spa{c}.{d}\spb{e}.{f}t_{efg}\spba{g}.{{ef}}.{d}}
H_2[ a,P_{fg}|e] ;b , c ;P_{ab}] \, ,
\cr}
\equn
$$
where,
$$
|X_{B1a}\ra = \spb{d}.{a}\spa{b}.{e}|a\ra + \spba{d}.{{bcd}}.{e} |b\ra \, .
\equn
$$
\bigskip
$$
\eqalign{&\hskip-1.0truecm
D_C[a,b,c,d,e,f,g]= 
\cr &\hskip-1.0truecm{1  \over  \spb{b}.{c}\spa{d}.{e}\spa{e}.{f}\spa{g}.{a}}G_5[ P_{ga}|b], f, P_{def}|c], P_{ga} P_{bc}|d\ra , g;     e, e, X_{C1a}, X_{C1a}, a, P_{ga}|c];     c; P_{bc};  P_{ga} ]\cr &\hskip-1.0truecm -{\spb{c}.{d}^3\spa{b}.{a}^2  \over  \spa{e}.{f}\spa{g}.{a}}     H_6[ f, g, P_{gab} P_{cd}|e\ra ,   P_{ef} P_{cd}|b\ra , P_{def}|d] , P_{def}|c] ;           a, b, e, e, e, e;         P_{ga}]    \cr &\hskip-1.0truecm-{1  \over  \spa{e}.{f}\spa{b}.{c}\spa{c}.{d}\spa{g}.{a}t_{bcd}}\cr \times & G_5[f,g,P_{ef} P_{cd}|b\ra , P_{ga} P_{bcd}|e\ra ,  P_{ga} P_{bcd}|d\ra ;  a, e, e, X_{C1c}, X_{C1c},  P_{ga} P_{bcd}|b\ra ; P_{bcd}|b];   P_{ef};     P_{ga} ] 
\cr &\hskip-1.0truecm-{\spa{b}.{e}^4\spa{a}.{g}^2\spb{g}.{f}^2  \over  \spa{b}.{c}\spa{c}.{d}\spa{d}.{e}\spa{g}.{a}\spba{f}.{{cde}}.{b}t_{fga}s_{ga}}     
\overline{H}_2[ a, P_{bcd}|e\ra; f, g; P_{ga}]
\cr &\hskip-1.0truecm+{\spba{f}.{{cd}}.{e}^4\spa{b}.{a}^2  \over  \spa{c}.{d}\spa{d}.{e}\spba{f}.{{de}}.{c}\spba{f}.{{ga}}.{b} t_{gab}t_{cde}\spa{g}.{a}}     H_2[g, P_{gab} P_{cd}|e\ra ; a, b;  P_{ga}]\cr &\hskip-1.0truecm-{\spb{f}.{d}^4\spa{b}.{a}^2  \over  \spa{b}.{c}\spb{d}.{e}\spb{e}.{f}\spba{f}.{{de}}.{c}\spa{g}.{a}t_{def}}H_2[ g, P_{ef}|d]; a, b; P_{ga}] \, ,
\cr}
\equn
$$
where,
$$
\eqalign{
|X_{C1a}\ra &  =  -\spa{e}.{a}( \spb{c}.{g}|g\ra +\spb{c}.{a}|{a}\ra)
                                    +\spb{c}.{b}\spa{b}.{e}|{a}\ra  \, ,
\cr
|X_{C1c}\ra &  =            \spa{e}.{b}t_{bcd}|a\ra
                                          +\spa{e}.{a}( \spab{b}.{{cd}}.{g}|{g}\ra
                                                         +\spab{b}.{{cd}}.{a}|{a}\ra)  \, .
\cr}
\equn
$$
\bigskip

\def\curlyfb{;}
\def\curlyb{;}
\bigskip
$$
\eqalign{
&\hskip-1.0truecm D_D[a,b,c,d,e,f,g]=
\cr &\hskip-1.0truecm
{1  \over  \spa{a}.{b}\spa{c}.{d}\spa{d}.{e}\spb{f}.{g}}
G_4[ b,  P_{ab}|g] , P_{ab} P_{fg}|e\ra ,P_{cde}|f] \curlyfb a,c,X_{D1},X_{D1}, P_{ab}|f] \curlyb f; P_{cde}; P_{ab} ]\cr &\hskip-1.0truecm-{ \spa{a}.{g}^2\spba{f}.{{cde}}.{c}^3   \over  \spa{a}.{b}\spa{c}.{d}\spa{d}.{e}\spaba{ e}.{ {cd}}.{ {ab}}.{ g}t_{cde}t_{gab} }
H_2[  b,  P_{cde}|f] ; a, g; P_{ab} ]\cr &\hskip-1.0truecm-{\spba{d}.{{ef}}.{g}^3\spa{a}.{g}^2  \over  \spa{a}.{b}\spa{e}.{f}\spa{f}.{g}\spb{c}.{d}\spba{c}.{{ab}}.{g}\spaba{ g}.{ {ab}}.{{cd}}.{e} }
H_2[ b , P_{cd} P_{ef}|g\ra ;  a,g; P_{ab}]  \cr &\hskip-1.0truecm+{1  \over  \spa{a}.{b}\spa{e}.{f}\spa{f}.{g}\spa{c}.{d}t_{efg} }\cr &\times
G_4[  b,  P_{ab} P_{efg}|e\ra , P_{ab} P_{efg}|d\ra ,  P_{cd} P_{ef}|g\ra  \curlyfb a,c,    P_{ab} P_{ef}|g\ra ,  X_{D4}, X_{D4} \curlyb P_{ef}|g\ra ; P_ {cd}; P_{ab} ]\cr &\hskip-1.0truecm-{ \spba{b}.{{def}}.{g}^2  \over  \spb{a}.{b}\spa{d}.{e}\spa{e}.{f}\spa{f}.{g}\spba{ c}.{{abc}}.{g}t_{abc} }
\overline{H}_3[ a, P_{abc}|d\ra  ,  c; b, P_{abc}|g\ra , P_{abc}|g\ra ; P_{ab}   ] \, ,
\cr}
\equn
$$
where,
$$
\eqalign{
|X_{D1}\ra &=-|{a}\ra\spba{f}.{{ga}}.{c}+|{b}\ra\spb{b}.{f}\spa{a}.{c} \, ,  
\cr
|X_{D4}\ra &=|{a}\ra\spba{d}.{{ef}}.{g}\spa{c}.{d}+|{c}\ra\spba{b}.{{ef}}.{g}\spa{a}.{b} \, .  
\cr}
\equn
$$

$$
\eqalign{ &\hskip-1.0truecm
D_E[a,b,c,d,e,f,g]=
\cr &\hskip-1.0truecm
-{1\over \spa{c}.{d}\spa{e}.{f}\spa{f}.{g} s_{ab}}\cr &\hskip-1.0truecm\times \left(\begin{array}{cc} \spba{f}.{{eb}}.{d}^2& G_5[  a , g, c, P_{fg}|e] ,  P_{cd}|e] \curlyfb d,e, b,f,b,f \curlyb a ;  P_{efg}; P_{ab}] \\+\spba{g}.{{eb}}.{d}^2& G_5[  a , g, c, P_{fg}|e] ,  P_{cd}|e] \curlyfb d,e, b,g,b,g \curlyb a ;  P_{efg}; P_{ab}]\\+(\spb{f}.{a}\spa{b}.{d})^2& G_5[  a , g, c, P_{fg}|e] ,  P_{cd}|e] \curlyfb d,e, a,f,a,f \curlyb a ;  P_{efg}; P_{ab}]\\+(\spb{g}.{a}\spa{b}.{d})^2& G_5[  a , g, c, P_{fg}|e] ,  P_{cd}|e] \curlyfb d,e, a,g,a,g \curlyb a ;  P_{efg}; P_{ab}]\\+2\spba{f}.{{eb}}.{d}\spba{g}.{{eb}}.{d}& G_5[  a , g, c, P_{fg}|e] ,  P_{cd}|e] \curlyfb d,e, b,f,b,g \curlyb a ;  P_{efg}; P_{ab}]\\+2\spba{f}.{{eb}}.{d}\spb{f}.{a}\spa{b}.{d}& G_5[  a , g, c, P_{fg}|e] ,  P_{cd}|e] \curlyfb d,e, b,f,a,f \curlyb a ;  P_{efg}; P_{ab}]\\+2\spba{f}.{{eb}}.{d}\spb{g}.{a}\spa{b}.{d}& G_5[  a , g, c, P_{fg}|e] ,  P_{cd}|e] \curlyfb d,e, b,f,a,g \curlyb a ;  P_{efg}; P_{ab}]\\+2\spba{g}.{{eb}}.{d}\spb{f}.{a}\spa{b}.{d}& G_5[  a , g, c, P_{fg}|e] ,  P_{cd}|e] \curlyfb d,e, b,g,a,f \curlyb a ;  P_{efg}; P_{ab}]\\+2\spba{g}.{{eb}}.{d}\spb{g}.{a}\spa{b}.{d}& G_5[  a , g, c, P_{fg}|e] ,  P_{cd}|e] \curlyfb d,e, b,g,a,g \curlyb a ;  P_{efg}; P_{ab}]\\+2\spb{f}.{a}\spa{b}.{d}\spb{g}.{a}\spa{b}.{d}& G_5[  a , g, c, P_{fg}|e] ,  P_{cd}|e] \curlyfb d,e, a,f,a,g \curlyb a ;  P_{efg}; P_{ab}]\end{array}\right)\cr &\hskip-1.0truecm-{\spa{e}.{d}\over \spa{c}.{d}\spa{e}.{f}\spa{f}.{g}\spa{d}.{f}s_{ab}}\cr &\hskip-1.0truecm\times\left(\begin{array}{cc} \spba{f}.{{eb}}.{d}^2& G_5[  a , g, c,  P_{fg}|e] ,  P_{cd}|e] \curlyfb d,f,b,f,b,f \curlyb a ; P_{fgX_{E3}} ; P_{ab}]   \\+(\spba{g}.{{eb}}.{d})^2& G_5[  a , g, c,  P_{fg}|e] ,  P_{cd}|e] \curlyfb d,f,b,g,b,g \curlyb a ; P_{fgX_{E3}} ; P_{ab}]   \\+(\spb{f}.{a}\spa{b}.{d})^2& G_5[  a , g, c,  P_{fg}|e] ,  P_{cd}|e] \curlyfb d,a,f,f,a,f \curlyb a ; P_{fgX_{E3}} ; P_{ab}]   \\+(\spb{g}.{a}\spa{b}.{d})^2& G_5[  a , g, c,  P_{fg}|e] ,  P_{cd}|e] \curlyfb d,f,a,g,a,g \curlyb a ; P_{fgX_{E3}} ; P_{ab}]   \\+2\spba{f}.{{eb}}.{d}\spba{g}.{{eb}}.{d}& G_5[  a , g, c,  P_{fg}|e] ,  P_{cd}|e] \curlyfb d,f,b,f,b,g \curlyb a ; P_{fgX_{E3}} ; P_{ab}]   \\+2\spba{f}.{{eb}}.{d}\spb{f}.{a}\spa{b}.{d}& G_5[  a , g, c,  P_{fg}|e] ,  P_{cd}|e] \curlyfb d,f,b,f,a,f \curlyb a ; P_{fgX_{E3}} ; P_{ab}]   \\+2\spba{f}.{{eb}}.{d}\spb{g}.{a}\spa{b}.{d}& G_5[  a , g, c,  P_{fg}|e] ,  P_{cd}|e] \curlyfb d,f,b,f,a,g \curlyb a ; P_{fgX_{E3}} ; P_{ab}]   \\+2\spba{g}.{{eb}}.{d}\spb{f}.{a}\spa{b}.{d}& G_5[  a , g, c,  P_{fg}|e] ,  P_{cd}|e] \curlyfb d,f,b,g,a,f \curlyb a ; P_{fgX_{E3}} ; P_{ab}]   \\+2\spba{g}.{{eb}}.{d}\spb{g}.{a}\spa{b}.{d}& G_5[  a , g, c,  P_{fg}|e] ,  P_{cd}|e] \curlyfb d,f,b,g,a,g \curlyb a ; P_{fgX_{E3}} ; P_{ab}]   \\+2\spb{f}.{a}\spa{b}.{d}\spb{g}.{a}\spa{b}.{d}& G_5[  a , g, c,  P_{fg}|e] ,  P_{cd}|e] \curlyfb d,f,a,f,a,g \curlyb a ; P_{fgX_{E3}} ; P_{ab}]   \end{array}\right)\cr &\hskip-1.0truecm+{\spa{d}.{e}^3\over \spa{a}.{b}\spa{e}.{f}\spa{f}.{g}\spa{d}.{f}\spba{c}.{{de}}.{f} }G_4[ g,a, P_{ab} P_{cde}|f\ra ,P_{abc}P_{de}|f\ra  \curlyfb b,f,f,X_{E1b},X_{E1b}\curlyb c ; P_{fgX_{E3}} ;P_{ab} ] \cr &\hskip-1.0truecm+{ \spba{c}.{{cde}}.{b}^2 \over\spa{f}.{g}\spa{a}.{b}\spb{c}.{d}\spb{d}.{e}\spba{c}.{{de}}.{f} t_{cde} }
H_3[ a, g , P_{cd}|e] ; b ,P_{de}|c] ,P_{de}|c] ; P_{ab} ]\cr &\hskip-1.0truecm-{  \spa{d}.{e}^4\spb{g}.{a}^2 \over\spa{c}.{d}\spa{d}.{e}\spa{e}.{f}\spb{a}.{b}\spba{g}.{{ab}}.{c} t_{gab} }H_2[ a , P_{ab} P_{gab}|f\ra ; b,  P_{ab}|g] ; P_{ab} ]\cr &\hskip-1.0truecm -{ \spa{d}.{e}^3\spba{g}.{{def}}.{b}^2 \over\spa{a}.{b}\spa{e}.{f}\spba{g}.{{def}}.{c}\spba{g}.{{ef}}.{d} t_{abc} t_{def} }
H_3 [  a, c ,  P_{abc} P_{de}|f\ra ; b,   P_{def}|g] , P_{def}|g]  ; P_{ab} ] \cr &\hskip-1.0truecm-{\spb{f}.{g}^3\spa{b}.{d}^2 \over \spa{a}.{b}\spa{c}.{d}\spb{e}.{f}\spba{g}.{{ef}}.{d} t_{efg} }
H_3[  a , c , P_{fg}|e] ;  b,d,d ; P_{ab} ] \, ,
\cr}
\equn
$$
where,
$$
\eqalign{
|X_{E1b}\ra & =\spb{c}.{g}\spa{g}.{f} |b\ra +\spb{c}.{a}\spa{a}.{b}|f\ra \, ,
\cr
\Slash{k}_{X_{E3}} & ={\spa{e}.{d}\over \spa{f}.{d}} |f\ra \la e| \, .
\cr}
\equn
$$

$$
\small
\eqalign{
&\hskip-1.0truecm D_F[a,b,c,d,e,f,g]=
\cr &\hskip-1.0truecm{1  \over  \spa{a}.{b}\spb{c}.{d}\spa{e}.{f}\spa{f}.{g}}G_4[ b, g, P_{efg}|d] , P_{ab} P_{cd}|e\ra  \curlyfb     f, f, a, X_{F1a}, X_{F1a} \curlyb     c;  P_{cd} ; P_{ab}] \cr &\hskip-1.0truecm-{\spb{e}.{c}  \over  \spa{f}.{g}\spb{d}.{e}\spa{a}.{b}}H_7[b, g, c, P_{fg}|e], P_{fg} P_{de}|c\ra, P_{abc}|d],  P_{abc} P_{de}|f\ra ;    a, c, f, f,         P_{abc}|e],        X_{F1b},         X_{F1b}                 ;    P_{ab} ] \cr &\hskip-1.0truecm+{1  \over  \spa{f}.{g}\spb{d}.{e}\spa{a}.{b}}H_7[b, c, g, P_{fg} P_{de}|c\ra, P_{abc} P_{de}|f\ra, P_{abc}|d], P_{fg}|e] ;    a, f, f, P_{abc}|e],         P_{ab}|e],           X_{F1b},        X_{F1b}       ;   P_{ab}] \cr &\hskip-1.0truecm-{1  \over  \spa{f}.{g}\spa{a}.{b}\spa{c}.{d}\spa{d}.{e}t_{cde}}\cr &\times 
G_5[ b, g, P_{fg} P_{de}|c\ra, P_{ab} P_{cd}|e\ra, P_{ab} P_{cde}|f\ra \curlyfb     a, f, f, P_{ab} P_{cde}|d\ra, X_{F1c}, X_{F1c} \curlyb    P_{cde}|d]; P_{fg}; P_{ab}]\cr &\hskip-1.0truecm-{ \spa{d}.{f}^4\spb{g}.{b}^2\spa{b}.{a}  \over  \spa{c}.{d}\spa{d}.{e}\spa{e}.{f}\spba{g}.{{def}}.{c}s_{ab}t_{gab} } \overline{H}_2[ P_{ab}|b\ra , P_{cde}|f\ra ; g,  P_{ab}|a\ra ; P_{ab} ] \cr &\hskip-1.0truecm-{ \spa{d}.{f}^4\spab{g}.{{abc}}.{a}^2  \over  \spa{a}.{b}\spa{d}.{e}\spa{e}.{f}\spba{g}.{{abc}}.{c}\spba{g}.{{abc}}.{d}t_{abc}t_{def}}H_3[b, c, P_{abc} P_{de}|f\ra ;  P_{abc}|g],  P_{abc}|g],a  ;  P_{ab}]\cr &\hskip-1.0truecm-{ \spb{g}.{e}^4\spa{d}.{a}^2  \over \spa{a}.{b}\spa{c}.{d}\spb{e}.{f}\spb{f}.{g}\spba{g}.{{efg}}.{d}t_{efg}}H_3[b, c, P_{efg}|e] ;  d, d, a  ;  P_{ab}] \, ,
\cr}
\equn
$$
where,
$$
\eqalign{
|X_{F1a}\ra & = |{f}\ra\spb{c}.{b}\spa{b}.{a}
                                           +|{a}\ra\spba{c}.{{efg}}.{f}\, ,
\cr
|X_{F1b}\ra & = |{f}\ra \spb{e}.{d}\spa{d}.{a}
                                           +\spa{f}.{a}( |{f}\ra\spb{e}.{f}+|{g}\ra\spb{e}.{g}) \, ,
\cr
|X_{F1c}\ra & = |{a}\ra\spa{f}.{g}\spba{g}.{{cde}}.{d}
                                            +|{f}\ra\spa{a}.{b}\spba{b}.{{cde}}.{d} \, .
\cr}
\equn
$$

\end{document}